\documentclass{article}

\usepackage{arxiv}

\usepackage[utf8]{inputenc} 
\usepackage[T1]{fontenc}    
\usepackage[hidelinks]{hyperref}       
\usepackage{url}            
\usepackage{booktabs}       
\usepackage{amsfonts}       
\usepackage{nicefrac}       
\usepackage{microtype}      
\usepackage{lipsum}
\usepackage[pdftex]{graphicx}
\usepackage{bookmark}

\usepackage{algorithmic}
\usepackage[ruled]{algorithm2e}

\usepackage[xindy]{glossaries} 

\newacronym{acaces}{ACACES}{Advanced Computer Architecture and Compilation for high-performance Embedded Systems}
\newacronym{acm}{ACM}{Association for Computing Machinery}
\newacronym{aig}{AIG}{AND-inverter graph}
\newacronym{ai}{AI}{artificial intelligence}
\newacronym{ann}{ANN}{artifical neural network}
\newacronym{asic}{ASIC}{application-specific integrated circuit}
\newacronym{asr}{ASR}{automatic speech recognition}
\newacronym{avx}{AVX}{advanced vector extensions}
\newacronym{aws}{AWS}{Amazon Web Services}
\newacronym{bat}{BAT}{Baidu, Alibaba, Tencent}
\newacronym{bert}{BERT}{bidirectional encoder representations from transformers}
\newacronym{bfloat16}{bfloat16}{brain floating point 16-bit}
\newacronym{bic}{BIC}{bit clear}
\newacronym{blas}{BLAS}{basic linear algebra subprograms}
\newacronym{bnn}{BNN}{binary neural network}
\newacronym{bram}{BRAM}{block RAM}
\newacronym{brnn}{BRNN}{bidirectional recurrent neural network}
\newacronym{cal}{CAL}{Computer Architecture Letters}
\newacronym{cifar}{CIFAR}{Canadian Institute For Advanced Research}
\newacronym{clb}{CLB}{configurable logic block}
\newacronym{cnn}{CNN}{convolutional neural network}
\newacronym{cpu}{CPU}{central processing unit}
\newacronym{cudnn}{cnDNN}{NVIDIA CUDA Deep Neural Network}
\newacronym{dag}{DAG}{directed acyclic graph}
\newacronym{darpa}{DARPA}{Defense Advanced Research Projects Agency}
\newacronym{dbn}{DBN}{deep belief network}
\newacronym{des}{DES}{data encryption standard}
\newacronym{dl}{DL}{deep learning}
\newacronym{dnn}{DNN}{deep neural network}
\newacronym{dram}{DRAM}{dynamic RAM}
\newacronym{dsp}{DSP}{digital signal processor}
\newacronym{dvs}{DVS}{dynamic vision sensor}
\newacronym{eda}{EDA}{electronic design automation}
\newacronym{edram}{eDRAM}{embedded \gls{dram}}
\newacronym{eie}{EIE}{efficient inference engine}
\newacronym{elu}{ELU}{exponential linear unit}
\newacronym{fc}{FC}{fully connected}
\newacronym{fft}{FFT}{fast fourier transform}
\newacronym{fifo}{FIFO}{first in first out}
\newacronym{flopoco}{FloPoCo}{FLOating-POint COres}
\newacronym{flops}{FLOPS}{floating-point operations per seconds}
\newacronym{flop}{FLOP}{floating-point operations per seconds}
\newacronym{fma}{FMA}{fused multiply add}
\newacronym{fpga}{FPGA}{field programmable gate array}
\newacronym{fps}{FPS}{frames per second}
\newacronym{fpu}{FPU}{floating-point unit}
\newacronym{fp}{FP}{floating-point}
\newacronym{fsm}{FSM}{finite-state machine}
\newacronym{gan}{GAN}{generative adversarial network}
\newacronym{gdpr}{GDPR}{General Data Protection Regulation}
\newacronym{gemm}{GEMM}{general matrix multiply}
\newacronym{gflops}{GFLOPS}{giga \gls{flops}}
\newacronym{gmafia}{GMAFIA}{Google, Microsoft, Apple, Facebook, Intel, Amazon}
\newacronym{gops}{GOPS}{giga operations per second}
\newacronym{gpgpu}{GPGPU}{general purpose \gls{gpu}}
\newacronym{gpu}{GPU}{graphics processor unit}
\newacronym{hdl}{HDL}{hardware description language}
\newacronym{hipeac}{HiPEAC}{High Performance and Embedded Architecture and Compilation}
\newacronym{hls}{HLS}{high-level synthesis}
\newacronym{hobflops}{HOBFLOPS}{hardware optimized bitslice-parallel floating-point operators}
\newacronym{hpm}{HPM}{high performance mobile}
\newacronym{ibm}{IBM}{International Business Machines}
\newacronym{ic}{IC}{integrated circuit}
\newacronym{ieee}{IEEE}{Institute of Electrical and Electronic Engineers}
\newacronym{iet}{IET}{Institution of Engineering and Technology}
\newacronym{ifm}{IFM}{input feature map}
\newacronym{ilp}{ILP}{instruction-level parallelism}
\newacronym{ilsvrc}{ILSVRC}{ImageNet large scale visual recognition challenge}
\newacronym{iob}{IOB}{input output buffer}
\newacronym{iot}{IoT}{internet of things}
\newacronym{ip}{IP}{intellectual property}
\newacronym{isa}{ISA}{instruction set architecture}
\newacronym{iso}{ISO}{International Organization for Standardization}
\newacronym{isscc}{ISSCC}{International Solid-State Circuits Conference}
\newacronym{jpeg}{JPEG}{Joint Photographic Experts Group}
\newacronym{knn}{KNN}{K-nearest neighbour}
\newacronym{lpddr}{LPDDR}{low-power double data rate}
\newacronym{lstm}{LSTM}{long short term memory}
\newacronym{lut}{LUT}{look up table}
\newacronym{lzc}{LZC}{leading zeros count}
\newacronym{lzoc}{LZOC}{Leading Zero or One Count}
\newacronym{mac}{MAC}{multiply-accumulate}
\newacronym{mcmk}{MCMK}{multi-channel, multi-kernel convolution}
\newacronym{mec}{MEC}{memory-efficient convolution}
\newacronym{mit}{MIT}{Massachusetts Institute of Technology}
\newacronym{mlp}{MLP}{multilayer perceptron}
\newacronym{ml}{ML}{machine learning}
\newacronym{mnist}{MNIST}{Modified National Institute of Standards and Technology}
\newacronym{mosfet}{MOSFET}{metal oxide semi-conductor field effect transistor}
\newacronym{mux}{MUX}{multiplexer}
\newacronym{nan}{NAN}{not-a-number}
\newacronym{ncs}{NCS}{neural compute stick}
\newacronym{nlp}{NLP}{natrual language processing}
\newacronym{nn}{NN}{neural network}
\newacronym{npu}{NPU}{network processing unit}
\newacronym{nre}{NRE}{non-recurring engineering}
\newacronym{ofm}{OFM}{output feature map}
\newacronym{ops}{OPS}{operations per second}
\newacronym{pasm}{PASM}{parallel accumulate shared MAC}
\newacronym{pas}{PAS}{parallel accumulate and store}
\newacronym{pcb}{PCB}{printed circuit board}
\newacronym{pcie}{PCIe}{peripheral component interconnect express}
\newacronym{pe}{PE}{processing element}
\newacronym{pflops}{PFLOPS}{peta \gls{flops}}
\newacronym{prelu}{PRELU}{parameteric rectified linear unit}
\newacronym{qcd}{QCD}{quantum chromodynamics}
\newacronym{ram}{RAM}{random access memory}
\newacronym{rbm}{RBM}{restricted Boltzmann machine}
\newacronym{relu}{ReLU}{rectified linear unit}
\newacronym{rl}{RL}{reinforcement learning}
\newacronym{rnn}{RNN}{recurrent neural network}
\newacronym{roi}{ROI}{return on investment}
\newacronym{ros}{ROS}{robot operating system}
\newacronym{rtl}{RTL}{register transfer logic}
\newacronym{sat}{SAT}{satisfiability}
\newacronym{sdc}{SDC}{Synopsys design constraint}
\newacronym{sfi}{SFI}{Science Foundation Ireland}
\newacronym{sgd}{SGD}{stochastic gradient descent}
\newacronym{shave}{SHAVE}{streaming hybrid architecture vector engine}
\newacronym{simd}{SIMD}{single instruction multiple data}
\newacronym{sipp}{SIPP}{streaming image processing pipeline}
\newacronym{slide}{SLIDE}{sub-linear deep learning engine}
\newacronym{snarc}{SNARC}{Stochastic Neural Analog Reinforcement Calculator}
\newacronym{soi}{SOI}{silicon on insulator}
\newacronym{sop}{SOP}{sum-of-products}
\newacronym{sram}{SRAM}{static RAM}
\newacronym{svhn}{SVHN}{street view house numbers}
\newacronym{svm}{SVM}{support vector machine}
\newacronym{swar}{SWAR}{\gls{simd} within a register}
\newacronym{taco}{TACO}{Transactions on Architecture and Code Optimization}
\newacronym{tcl}{TCL}{tool command language}
\newacronym{tco}{TCO}{total cost of ownership}
\newacronym{tf32}{TP32}{tensor float 32}
\newacronym{tflops}{TFLOPS}{tera \gls{flops}}
\newacronym{tnn}{TNN}{ternary neural network}
\newacronym{tops}{TOPS}{tera operations per second}
\newacronym{tpu}{TPU}{tensor processing unit}
\newacronym{tvlsi}{TVLSI}{Transactions on Very Large Scale Integration}
\newacronym{twn}{TWN}{ternary weight network}
\newacronym{vfp}{VFP}{vector floating point}
\newacronym{vhdl}{VHDL}{very high speed integrated circuits hardware description language}
\newacronym{vliw}{VLIW}{very long instruction word}
\newacronym{xdc}{XDC}{Xilinx design constraint}
\newacronym{xpe}{XPE}{Xilinx power estimator}

\usepackage[caption=false]{subfig}

\usepackage{listings}
\usepackage{color}

\definecolor{codegreen}{rgb}{0,0.6,0}
\definecolor{codegray}{rgb}{0.5,0.5,0.5}
\definecolor{codepurple}{rgb}{0.58,0,0.82}
\definecolor{backcolour}{rgb}{1,1,1}

\lstdefinestyle{mystyle}{
	backgroundcolor=\color{backcolour},   
	commentstyle=\color{blue},
	keywordstyle=\bfseries\color{blue},
	numberstyle=\tiny\color{codegray},
	stringstyle=\color{codepurple},
	basicstyle=\footnotesize,
	breakatwhitespace=false,         
	breaklines=true,                 
	captionpos=b,                    
	keepspaces=true,                 
	numbers=left,                    
	numbersep=5pt,                  
	showspaces=false,                
	showstringspaces=false,
	showtabs=false,                  
	tabsize=2
}

\usepackage{array}

\usepackage{url}

\newcommand{\etal}{\textit{et al}., }
\newcommand{\etalp}{\textit{et al.'s}, }
\newcommand{\ie}{\textit{i}.\textit{e}., }
\newcommand{\eg}{\textit{e}.\textit{g}., }

\title{HOBFLOPS CNNs: Hardware Optimized Bitslice-Parallel Floating-Point Operations for Convolutional Neural Networks}

\author{
  James Garland\thanks{
  This research is supported by Science Foundation Ireland, Project 12/IA/1381. We thank the Institute of Technology Carlow, Carlow, Ireland for their support.} \\
  The School of Computer Science and Statistics\\
  Trinity College Dublin, The University of Dublin\\
  College Green, Dublin 2, Ireland \\
  \texttt{jgarland@tcd.ie} \\
   \And
 David Gregg \\
  The School of Computer Science and Statistics\\
  Trinity College Dublin, The University of Dublin\\
  College Green, Dublin 2, Ireland \\
  \texttt{david.gregg@cs.tcd.ie} \\
}

\begin{document}
\maketitle

\begin{abstract}
  \Acrfullpl{cnn} are typically trained using 16- or 32-bit \acrfull{fp} and researchers show that low-precision \gls{fp} can be highly effective for inference. Low-precision \gls{fp} can be implemented in \gls{fpga} and \gls{asic} accelerators, but existing processors do not generally support custom precision \gls{fp}.

We propose \acrfull{hobflops}, a method of generating efficient custom-precision emulated bitslice-parallel software \gls{fp} arithmetic. We generate custom-precision \gls{fp} routines optimized using a hardware synthesis design flow to create circuits. We provide standard cell libraries matching the bitwise operations on the target microprocessor architecture, and a code-generator to translate the hardware circuits to bitslice software equivalents. We exploit bitslice parallelism to create a very wide (32--512 element) vectorized \gls{cnn} convolution.

\Gls{hobflops} \gls{mac} performance in \gls{cnn} convolution on Arm and Intel processors are compared to Berkeley's Soft\gls{fp}16 equivalent \gls{mac}. \Gls{hobflops}16 outperforms Soft\gls{fp}16 by $8\times$ on Intel AVX512. \gls{hobflops} offers arbitrary-precision \gls{fp} with custom range and precision \eg \gls{hobflops}9 performs at $6\times$ the performance of \gls{hobflops}16 on Arm Neon. \Gls{hobflops} allows researchers to prototype different levels of custom \gls{fp} precision in the arithmetic of software \gls{cnn} accelerators. Furthermore, \gls{hobflops} fast custom-precision \gls{fp} \glspl{cnn} may be valuable in cases where memory bandwidth is limited.

\end{abstract}

\keywords{
Bitslice parallel arithmetic, datapath circuits, hardware accelerators, reduced floating-point precision arithmetic, convolutional neural networks.
}

{\section{Introduction}\label{sec:introduction}}
Many researchers have shown that \gls{cnn} inference is possible with low-precision integer \cite{TPUPerformance2017:Jouppi} and \acrfull{fp} \cite{ProjectBrainwave2018:Chung,Brainwave2018:Fowers} arithmetic. Almost all processors provide excellent support for 8-bit integer values, but not for bit-level custom precision \gls{fp} types, such as 9-bit \gls{fp}. Typically processors support a small number of relatively high-precision \gls{fp} types, such as 32- and 64-bit. However, there are good reasons why we might want to implement custom-precision \gls{fp} on regular processors. Researchers and hardware developers may want to prototype different levels of custom \gls{fp} precision that might be used for arithmetic in \gls{cnn} accelerators. Furthermore, fast custom-precision \gls{fp} \glspl{cnn} in software may be valuable in its own right, particularly in cases where memory bandwidth is limited.

While the choice of custom \gls{fp} in general purpose \glspl{cpu} is limited, \gls{fp} simulators, such as the excellent Berkeley's Soft\gls{fp} \cite{SoftFloat2002:Hauser}, are available. These simulators support certain ranges of custom \gls{fp} such as 16-, 32-, 64-, 80- and 128-bit with corresponding fixed-width mantissa and exponents.

We propose \acrfull{hobflops} which offers arbitrary-precision \gls{fp} arithmetic, using software \textit{bitslice parallel} \cite{BitsliceVectors2017:Xu} arithmetic, emulating any required precision \gls{fp} arithmetic at any bit-width of mantissa or exponent. Our goal is to use bit-slice packing to pack the vector registers of the microprocessor efficiently. Also, we exploit the bitwise logic optimization strategies of a commercial hardware synthesis tool to optimize the associated bitwise arithmetic that forms a multiplier and adder. We generate efficient arbitrary-precision software \gls{fp} emulation types and arithmetic, optimized using hardware tools and converted to the target processor's bitwise logic operators.

Existing methods of emulating \gls{fp} arithmetic in software primarily use existing integer instructions to implement the steps of \gls{fp} computation. This can work well for large, regular-sized \gls{fp} types such as \gls{fp}16, \gls{fp}32, or \gls{fp}64. Berkeley offer Soft\gls{fp} emulation \cite{SoftFloat2002:Hauser} for use where, for example, only integer precision instructions are available. Soft\gls{fp} emulation supports 16- to 128-bit arithmetic and does not support low bit-width custom precision \gls{fp} arithmetic or parallel arithmetic. \Gls{hobflops} offers fine grained customizable precision mantissa and exponent \gls{fp} bitwise arithmetic computed in parallel. To evaluate performance we benchmark \gls{hobflops}16 parallel \glspl{mac} against Berkeley's Soft\gls{fp}16 \textit{MulAdd}, in \gls{cnn} convolution implemented with Arm and Intel scalar and vector bitwise instructions. We show \gls{hobflops} offers significant performance boosts compared to Soft\gls{fp}. We then evaluate \gls{hobflops}8--\gls{hobflops}16e parallel arbitrary-precision performance. We argue that our software bitslice parallel \gls{fp} is both more efficient and offers greater bit-level customizability than conventional software \gls{fp} emulation.

\noindent We make the following contributions:
\begin{itemize}
  \item We present a full design flow from a VHDL \gls{fp} core generator to arbitrary-precision software bitslice parallel \gls{fp} operators, optimized using hardware design tools with our logic cell libraries, and our domain-specific code generator.
  \item We demonstrate how 3-input Arm NEON bitwise instructions \eg SEL (multiplexer) and AVX512 bitwise ternary operations can be used in standard cell libraries to improve the efficiency of the generated code significantly.
  \item We present an algorithm for implementing \gls{cnn} convolution with the very wide vectors that arise in bitslice parallel vector arithmetic.
  \item We evaluate \gls{hobflops}16 on Arm Neon and Intel AVX2 and AVX512 processors and find \gls{hobflops} achieves approximately $0.5\times$, $2.5\times$, and $8\times$ the performance of Berkeley Soft\gls{fp}16 respectively.
  \item We evaluate various widths of \gls{hobflops} from \gls{hobflops}8--\gls{hobflops}16e and find \eg \gls{hobflops}9 performs at approximately 45 million \glspl{mac}/second on Arm Neon processor around $6\times$ that of \gls{hobflops}16, and 2 billion \glspl{mac}/second on an Intel AVX512 platform, around $5\times$ that of \gls{hobflops}16.
  The increased performance is due to:
    \begin{itemize}
      \item Bitslice parallelism of the very wide vectorization of the \glspl{mac} of the \gls{cnn};
      \item Our efficient code generation flow.
    \end{itemize}
\end{itemize}

\noindent The rest of this article is organized as follows. Section~\ref{sec:background} gives background on other \gls{cnn} accelerators use of low-precision arithmetic types. Section~\ref{sec:approach} outlines bitslice parallel operations and introduces \gls{hobflops}, shows the design flow, types supported and how to implement arbitrary-precision \gls{hobflops} \gls{fp} arithmetic in a convolution layer of a \gls{cnn}. Section~\ref{sec:evaluation} shows results of our comparisons of \gls{hobflops}16 to Berkeley Soft\gls{fp}16 on Intel's AVX2 and AVX512 processors and show significant increases in performance. We also show results for \gls{hobflops}8--\gls{hobflops}16e emulation implemented on Arm Neon, Intel AVX2 and AVX512 processors. We conclude with Section~\ref{sec:conclusion}.

\section{Background}\label{sec:background}
Reduced-precision \gls{cnn} inference, particularly weight data of \glspl{cnn}, reduces computational requirements due to memory accesses, which dominate energy consumption. Energy and area costs are also reduced in \glspl{asic} and \glspl{fpga} \cite{EfficientProcessingOfDNN2017:Sze}.

Kang \etal \cite{ShortFloat2018:Kang} investigate short, reduced \gls{fp} representations that do not support \glspl{nan} and infinities. They show that shortening the width of the exponent and mantissa reduces the computational complexity within the multiplier logic. They compare fixed point integer representations with varying widths up to 8-bits of their short \gls{fp} in various \glspl{cnn}, and show around a 1\% drop in classification accuracy, with more than 60\% reduction in \gls{asic} implementation area. Their work stops at the byte boundary, leaving other arbitrary ranges open to investigation.

For their Project Brainwave \cite{ProjectBrainwave2018:Chung,Brainwave2018:Fowers}, Microsoft proposes MS-FP8 and MS-FP9, which are 8-bit and 9-bit \gls{fp} arithmetic that they exploit in a quantized \gls{cnn} \cite{ProjectBrainwave2018:Chung,Brainwave2018:Fowers}. Microsoft alters the Minifloat 8-bit that follows the IEEE-754 specification (1-sign bit, 4-exponent bits, 3-mantissa bits) \cite{FpArithmeticStandard2019:IEEE} by creating MS-\gls{fp}8, of 1-sign bit, 5-exponent bits, and 2-mantissa bits. MS-\gls{fp}8 gives a larger representative range due to the extra exponent bit but lower precision than Minifloat, caused by the reduced mantissa. MS-\gls{fp}8 more than doubles the performance compared to 8-bit integer operations, with negligible accuracy loss compared to full float. To improve the precision, they propose MS-\gls{fp}9, which increases the mantissa to 3 bits and keeps the exponent at 5 bits. Their later work \cite{Brainwave2018:Fowers} uses a shared exponent with their proposed MS-\gls{fp}8 / MS-\gls{fp}9, \ie one exponent pair used for many mantissae, sharing the reduced mantissa multipliers, something this work does not investigate. Their work remains at 8- and 9-bit for \gls{fpga} implementation and leaves the research area open for other bit-precision and range investigation.

Rzayev \etalp Deep Recon work \cite{DeepRecon2017:Rzayev} analyzes the computation costs of \glspl{dnn} and proposes a reconfigurable architecture to efficiently utilize computation and storage resources, thus allowing \gls{dnn} acceleration. They pay particular attention to comparing the prediction error of three \glspl{cnn} with different fixed and \gls{fp} precision. They demonstrate that \gls{fp} precision on the three \glspl{cnn} is 1-bit more efficient than fixed bit-width. They also show that the 8-bit \gls{fp} is around $7\times$ more energy-efficient and approximately $6\times$ more area efficient than 8-bit fixed precision.

The Flex Float C++ library proposed by Tagliavini \etal \cite{FlexFloat2018:Tagliavini} offers alternative \gls{fp} formats with variable bit-width mantissa and exponents. They demonstrate their Flex Float with two novel \gls{fp} formats, \textit{binary8} and \textit{binary16alt}. Using these two new formats, they demonstrate very efficient use in algorithms such as \gls{knn} and \gls{cnn}. They do not explore arbitrary bit-precision or other optimization techniques on the proposed number formats. 

Other researchers investigate optimizing different representations of \gls{fp} arithmetic. Xu \etal \cite{BitsliceVectors2017:Xu} propose bitslice parallel arithmetic and present \gls{fp} calculations undertaken on a fixed point unit. Instead of storing vectors in the traditional sense of storing 17-bit vectors inefficiently in a 32-bit register, they instead store thirty-two 17-bit words transformed into bitslice parallel format and stored in memory. Xu \etal manually construct bitwise arithmetic routines to perform integer or \gls{fp} arithmetic, while the vectors remain in a bitslice parallel format. When coded in C/C++ and AVX2 \gls{simd} instructions, they demonstrate this approach is efficient for low-precision vectors, such as 9-bit or 11-bit arbitrary \gls{fp} types. Xu \etalp work manually optimizes the arithmetic leaving automating the process an interesting area for investigation.

Researchers investigate different bit precision and representations of the \gls{fp} number base. Google's \gls{tpu} \gls{asic} \cite{TPUPerformance2017:Jouppi} implements \gls{bfloat16} \cite{BFloat162019:Google}, a 16-bit truncated IEEE-754 \gls{fp} single-precision format. \Gls{bfloat16} preserves dynamic range of the 32-bit format due to the 8-bit exponent. The precision is reduced in the mantissa from IEEE's 24-bits down to 7-bits. \Gls{bfloat16} is not restricted to machine learning; Intel Nervana use it in their \gls{npu}, Intel Xeon processors support AVX512 \gls{bfloat16} extensions, Intel Altera's \glspl{fpga} use it, Arm's ARMv8.6-A architecture, and ROCm libraries use \gls{bfloat16}. This work only reduces the bit-width of the mantissa and does not investigate different exponent or mantissa bit-widths. Our work addresses these arbitrary exponent and mantissa bit-widths.

Nvidia has proposed the new \gls{tf32} number representation \cite{TF322021:Choquette}, which is a sign bit, 8-bit exponent and 10-bit mantissa. This 19-bit floating-point format is an input format which truncates the IEEE \gls{fp}32 mantissa to 10-bits. \Gls{tf32} still produces 32-bit floating-point results to move and store in the \gls{gpu}. Similar to Google, Nvidia does not investigate other bit-widths of \gls{fp} range and precision, something our work does.

\section{Approach}\label{sec:approach}
\begin{figure}
  \centering
  \includegraphics[width=0.9\linewidth]{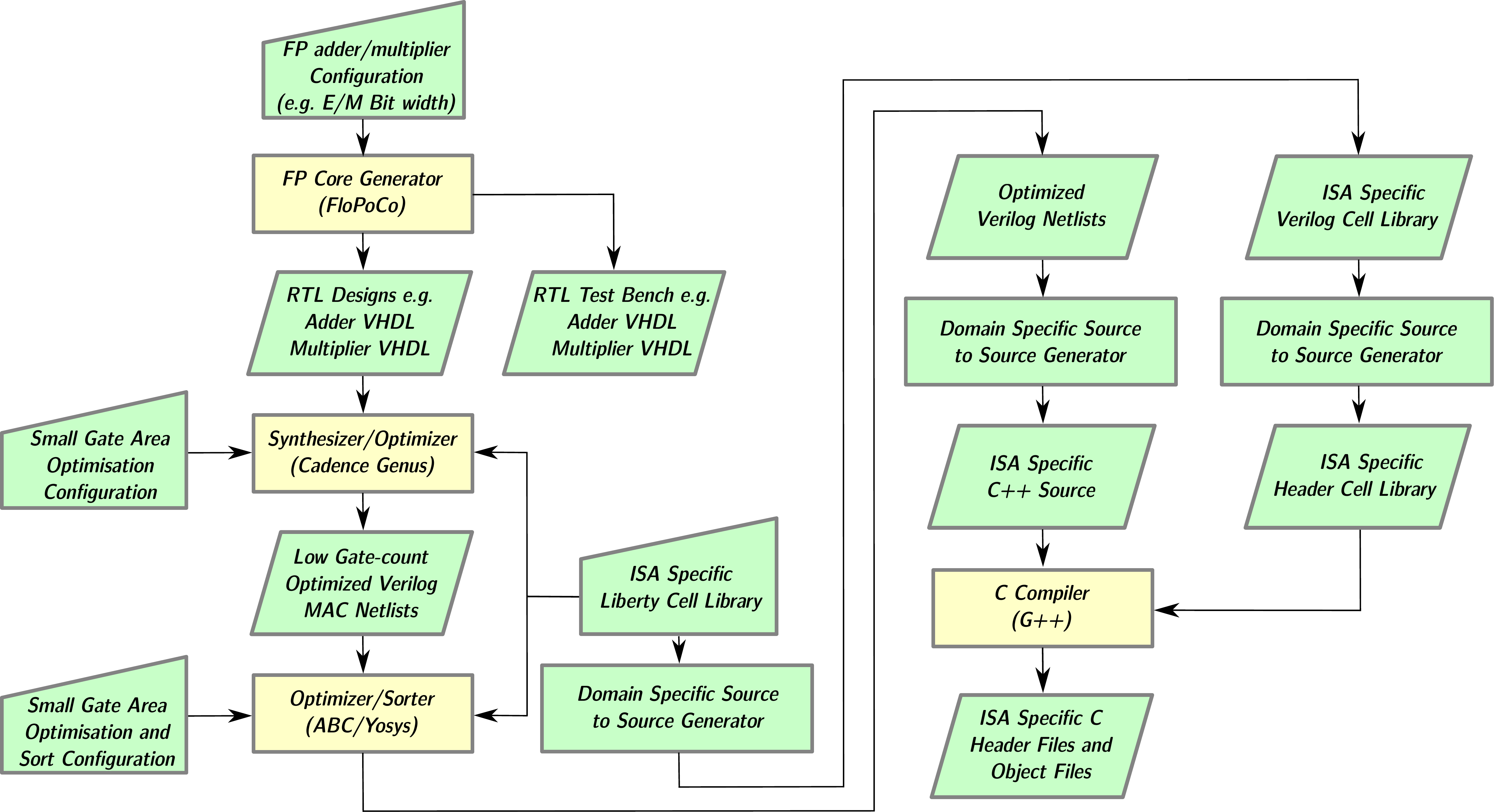}
  \caption{Flow for Creating HOBFLOPS Bitwise Operations. (Yellow signifies third party tools)}
  \label{fig:workflow}
\end{figure}

In this section, we present our approach to producing \gls{hobflops} arithmetic units that we demonstrate in a \gls{cnn} convolution layer. \Gls{hobflops} is a method for generating efficient software emulation parallel \gls{fp} arithmetic units optimized using hardware synthesis tools. \Gls{hobflops} investigates reduced complexity \gls{fp} \cite{ShortFloat2018:Kang} that is more efficient than fixed-point \cite{DeepRecon2017:Rzayev} by considering alternative \gls{fp} formats \cite{FlexFloat2018:Tagliavini} and register packing with bit-sliced arithmetic \cite{BitsliceVectors2017:Xu}.

Figure~\ref{fig:workflow} outlines our flow for creating \gls{hobflops} arithmetic units. We generate the \gls{rtl} representations of arbitrary-precision \gls{fp} multipliers and adders using the \gls{fp} unit generator, \gls{flopoco} \cite{GeneratingHighPerformanceCustomFpPipelines2009:Dinechin}. The floating-point accuracy of storage and computation of \gls{hobflops} adders and multipliers are configured by choosing the required \gls{fp} exponent and mantissa values of the VHDL produced by \gls{flopoco}.

We produce standard cell libraries of logic gate cells supported by the bitwise logic instructions of the target microprocessor architecture. For example, AND, OR, XOR, the SEL (multiplexer) bitwise instructions are supported on Arm Neon. The ternary logic \gls{lut} bitwise instructions of the Intel AVX512 are supported.

We use the \gls{asic} synthesizer tool, Cadence Genus in conjunction with our standard cell libraries and small gate area optimization and automation script to synthesize the adders and multipliers into Verilog netlists. The open-source synthesis and technology mapping suite, Yosys \gls{asic} synthesizer \cite{Yosys2013:Wolf} and ABC optimizer \cite{ABC2010:Brayton}, allows us to optimize further and topologically sort the netlists with the same cell libraries.

Our custom domain-specific source to source generator converts the topologically sorted netlists into a C/C++ header of bitwise operations. In parallel, we convert the hardware cell libraries into the equivalent C/C++ cell library headers of the target processor \gls{isa}. We create a \gls{cnn} convolution layer to include the \gls{hobflops} adder and multiplier and cell library headers corresponding to the target \gls{isa}, and compile with G++.

{\subsection{Arm Neon, Intel AVX2 and AVX512 Cell Libraries}
  \label{subsec:cellLibraries}}
\begin{table*}
\fontsize{8}{9}\selectfont
\centering
\caption{Cell Libraries' Support for Bitwise Logic Operations.}
\label{tab:gatesSupported}
\begin{tabular}{|ll|ll|ll|ll|ll|} 
\hline
\multicolumn{2}{|c|}{\begin{tabular}[c]{@{}c@{}} \textbf{Arm}\\\textbf{(64-bit)} \end{tabular}} &
\multicolumn{2}{c|}{\begin{tabular}[c]{@{}c@{}}\textbf{Arm Neon}\\\textbf{(128-bit) \if 0 \cite{ArmNeonIntrinsics2019:Arm}\fi} \end{tabular}} &
\multicolumn{2}{c|}{\begin{tabular}[c]{@{}c@{}}\textbf{Intel}\\\textbf{(64-bit)} \end{tabular}} &
\multicolumn{2}{c|}{\begin{tabular}[c]{@{}c@{}}\textbf{Intel AVX2}\\\textbf{(128-, 256-bit)} \end{tabular}} &
\multicolumn{2}{c|}{\begin{tabular}[c]{@{}c@{}}\textbf{Intel AVX512}\\\textbf{(512-bit) \if 0 \cite{IntrinsicsGuide2020:Intel}\fi} \end{tabular}}  \\ \hline
AND             & A \&{} B                        & AND & A \&{} B                                                               & AND                  & A \&{} B           & AND    & A \&{} B                  & LUT000                 & 0                                       \\
OR              & A \textbar{} B                  & OR  & A \textbar{} B                                                         & OR                   & A \textbar{} B     & OR     & A \textbar{} B            & LUT001                 & (A \textbar{} (B \textbar{} C)) \^{} 1  \\
XOR             & A \^{} B                        & XOR & A \^{} B                                                               & XOR                  & A \^{} B           & XOR    & A \^{} B                  & LUT002                 & \textasciitilde{} (B \textbar{} A) C    \\
NOT             & \textasciitilde{}A              & NOT & \textasciitilde{}A                                                     & NOT                  & \textasciitilde{}A & NOT    & \textasciitilde{}A        & LUT003                 & (B \textbar{} A) \^{} 1                 \\
ORN             & A \&{} (\textasciitilde{}B)     & ORN & A \textbar{} \textasciitilde{}B                                        &                      &                    & ANDNOT & \textasciitilde{}A \&{} B & LUT004                 & \textasciitilde{}(A \textbar{} C) B     \\
                &                                 & SEL & (\textasciitilde{}((S \&{} A) \textbar{} (\textasciitilde{}S \&{} B))) &                      &                    &        &                           & LUT005                 & (C \textbar{} A) \^{}1                  \\
                &                                 &     &                                                                        &                      &                    &        &                           & \textbf{\textit{...}}  & \textbf{\textit{... (truncated)}}       \\
                &                                 &     &                                                                        &                      &                    &        &                           & LUT253                 & A \textbar{} (B \textbar{} (C \^{} 1))  \\
                &                                 &     &                                                                        &                      &                    &        &                           & LUT254                 & A \textbar{} (B \textbar{} C)           \\
                &                                 &     &                                                                        &                      &                    &        &                           & LUT255                 & 1                                       \\
\hline
\end{tabular}
\end{table*}

We create three Synopsys Liberty standard cell libraries supporting the equivalent hardware gate-level representations of Arm Neon Intel X86\_64, AVX, AVX2 and AVX512 \gls{simd} vector intrinsics. The Arm Neon SEL (multiplexer) bitwise multiplexer instruction is a 3-input bitwise logic instruction, whereas all other Neon bitwise logic instructions modeled in the cell library are 2-input. The ternary logic \gls{lut} of the Intel AVX512 is a 3-input bitwise logic instruction that can implement 3-input boolean functions. An 8-bit immediate operand to this instruction specifies which of the 256 3-input functions should be used. We create all 256 equivalent cells in the Liberty cell library when targeting AVX512 devices. Table~\ref{tab:gatesSupported} lists the bitwise operations supported in the cell libraries for each architecture. The AVX512 column shows a truncated example subset of the available 256 bitwise logic instructions; see Intel's Intrinsics Guide and Software Developers manual for the complete set.

\begin{figure}
  \centering
  \subfloat[Full Adder Implemented in Intel AVX2 Intrinsic Bitwise Operations.]{\includegraphics[width=4.8cm]{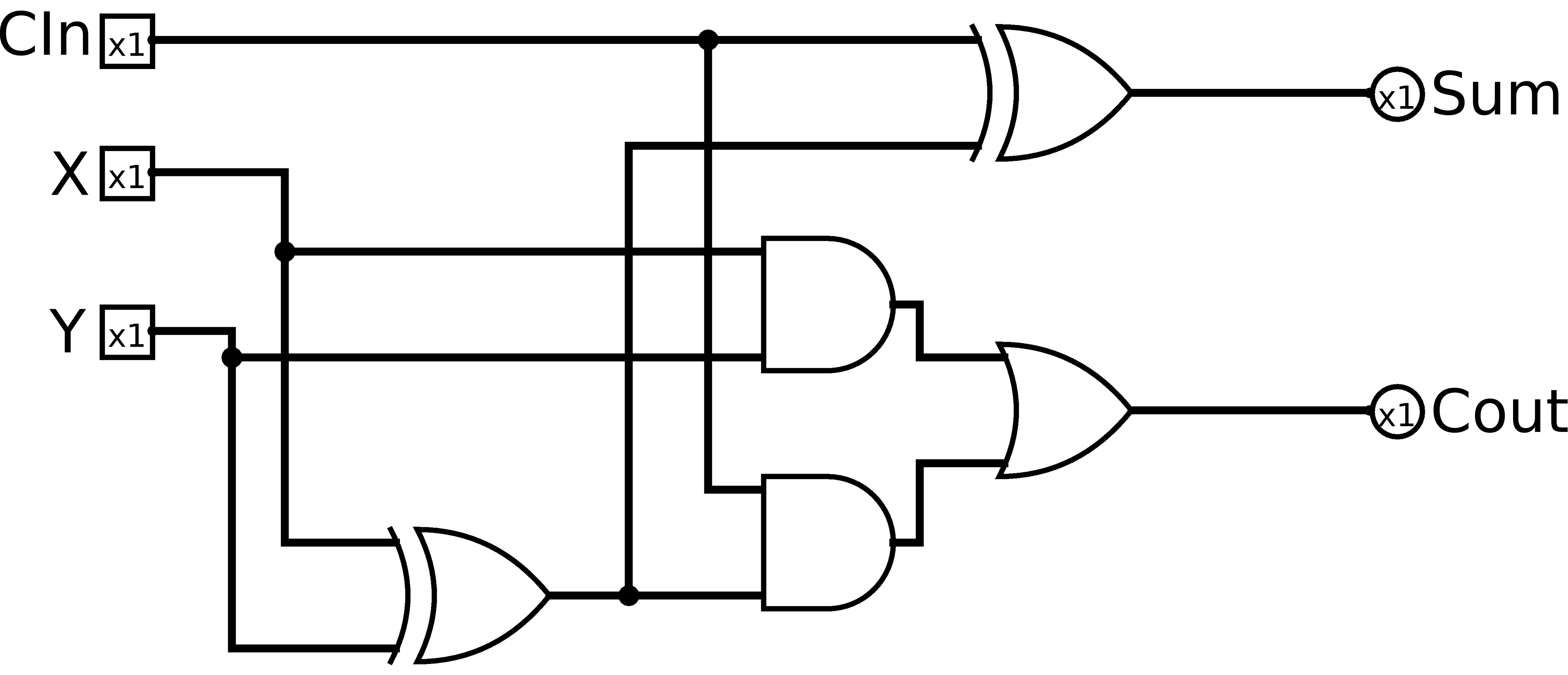}\label{fig:fullAdderAvx2}}\qquad
  \subfloat[Full Adder Implemented in Arm Neon Intrinsic Bitwise Operations.]{\includegraphics[width=4.8cm]{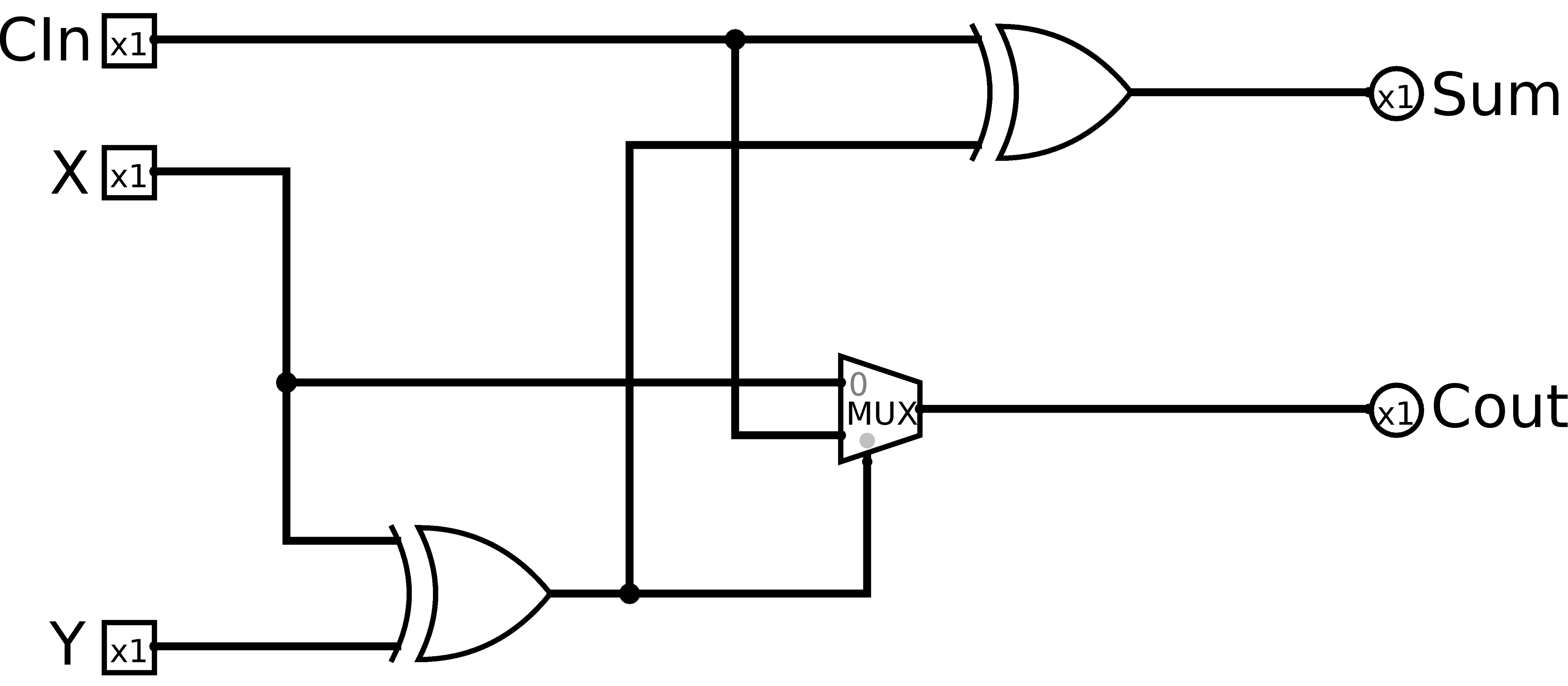}\label{fig:fullAdderNeon}}\qquad
  \subfloat[Full Adder Implemented in Intel AVX512 3-Input LUT Intrinsic Bitwise Operations.]{\includegraphics[width=4.8cm]{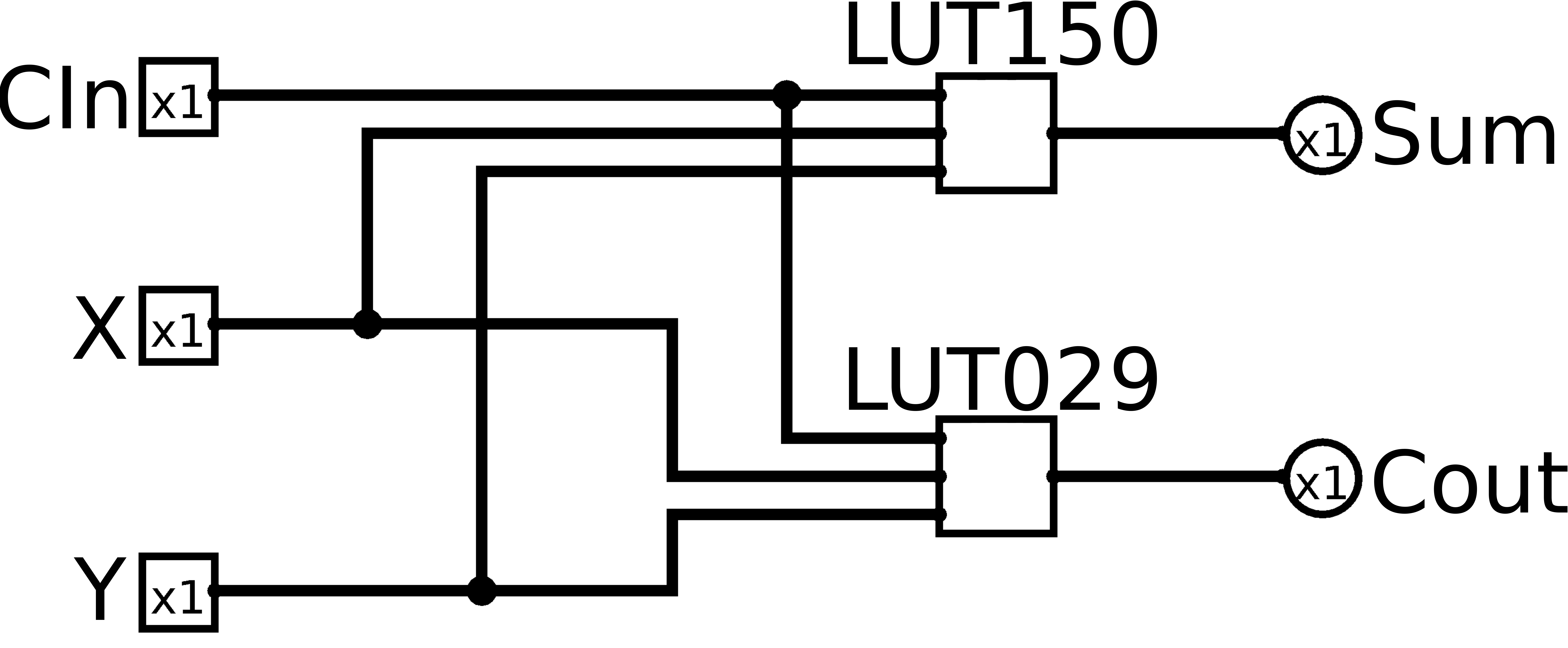}\label{fig:fullAdderAvx512}}
  \caption{Full Adders (Implemented in Intel's AVX2 and AVX512 and Arm's Neon Intrinsic Bitwise Operations.)}
  \label{fig:intelFullAdderspasmInOperation}
\end{figure}

\begin{lstlisting}[language={C++}, caption={Bitslice Parallel Adder for Two WIDTH-bit Wide Arrays of Unsigned Integers.}, label={lst:oneBitAdder},float]
#define WIDTH 2
void adder(u32 *x, u32 *y, u32 *sum, u32 *cout) {
  cout = 0;
  for (int i = 0; i < WIDTH; i++) {
    sum[i] = x[i] ^ y[i] ^ cout;
    cout = (cout & x[i] ^ y[i]) | (x[i] & y[i]);
  }
}
\end{lstlisting}

To demonstrate the capabilities of the cell libraries, we show an example of a single bit full adder. Figure~\ref{fig:fullAdderAvx2} shows a typical 5-gate full adder implemented with our AVX2 cell library. The C code function of Listing~\ref{lst:oneBitAdder} demonstrates these gates in code, if $WIDTH$ is set to $1$. The same full adder can be implemented in three Arm Neon bitwise logic instructions, Figure~\ref{fig:fullAdderNeon}, one of which is the SEL bitwise multiplexer instruction. Intel AVX512 intrinsics can implement the full adder in two 3-input bitwise ternary instructions, Figure~\ref{fig:fullAdderAvx512}. While the inputs and outputs of the hardware gate level circuits are single bits, these inputs and outputs are parallelized by the bit-width of the \gls{simd} vector registers. The bitwise instructions, when converted to bitwise software operations, produce extensive parallel scaling of the arithmetic.

{\subsection{Bitslice Parallel Operations}
  \label{subsec:bitSlicedOperations}}
\begin{figure}
  \centering
  \subfloat[Many 9-bit FP Values Transformed to 9-bit bitslice parallel FP Data Layout.]{\includegraphics[width=7cm]{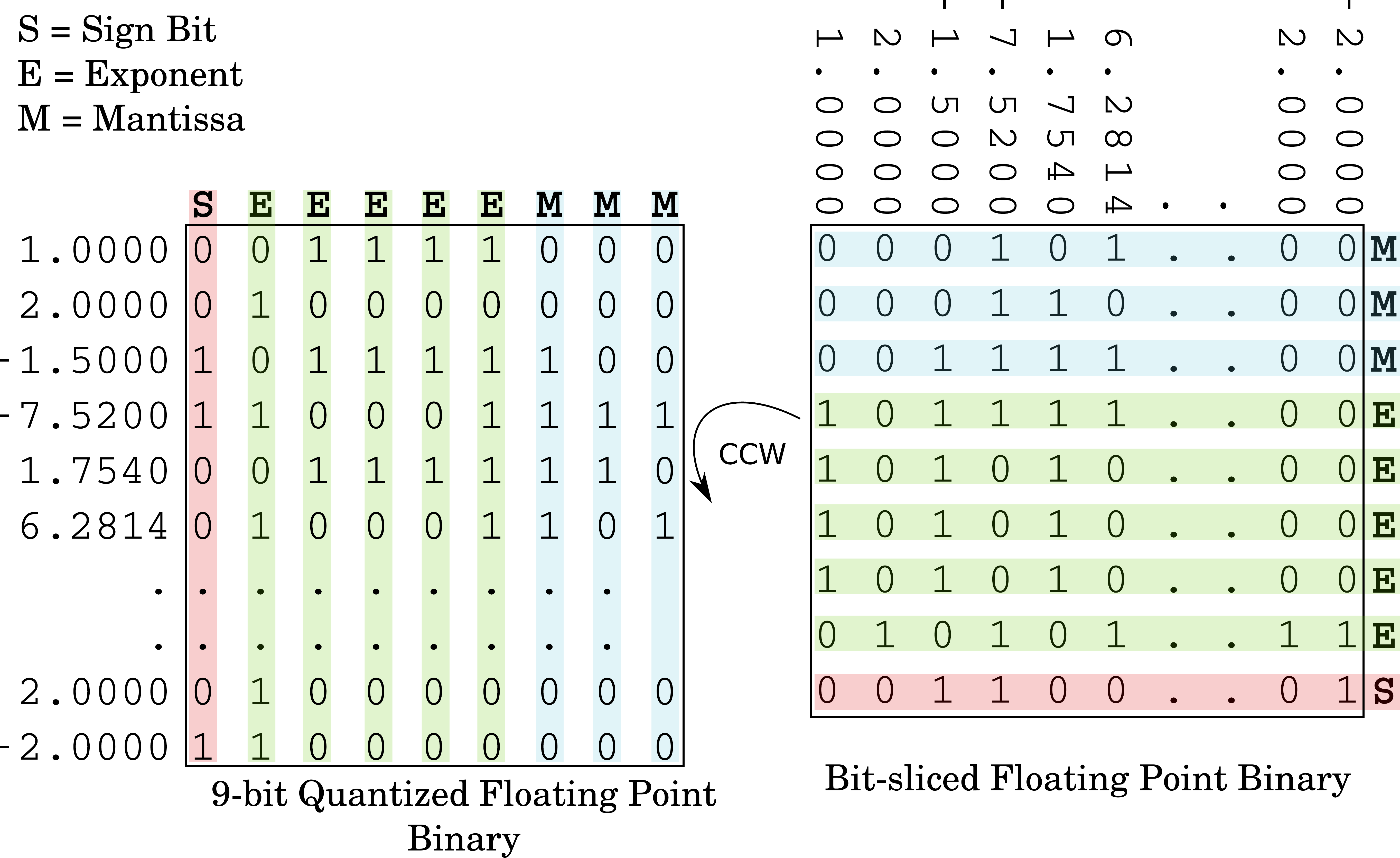}\label{fig:fp2bsfp}}\qquad
  \subfloat[Example 512 9-bit Bitslice Parallel FP Add Operation Using AVX512 Registers.]{\includegraphics[width=7cm]{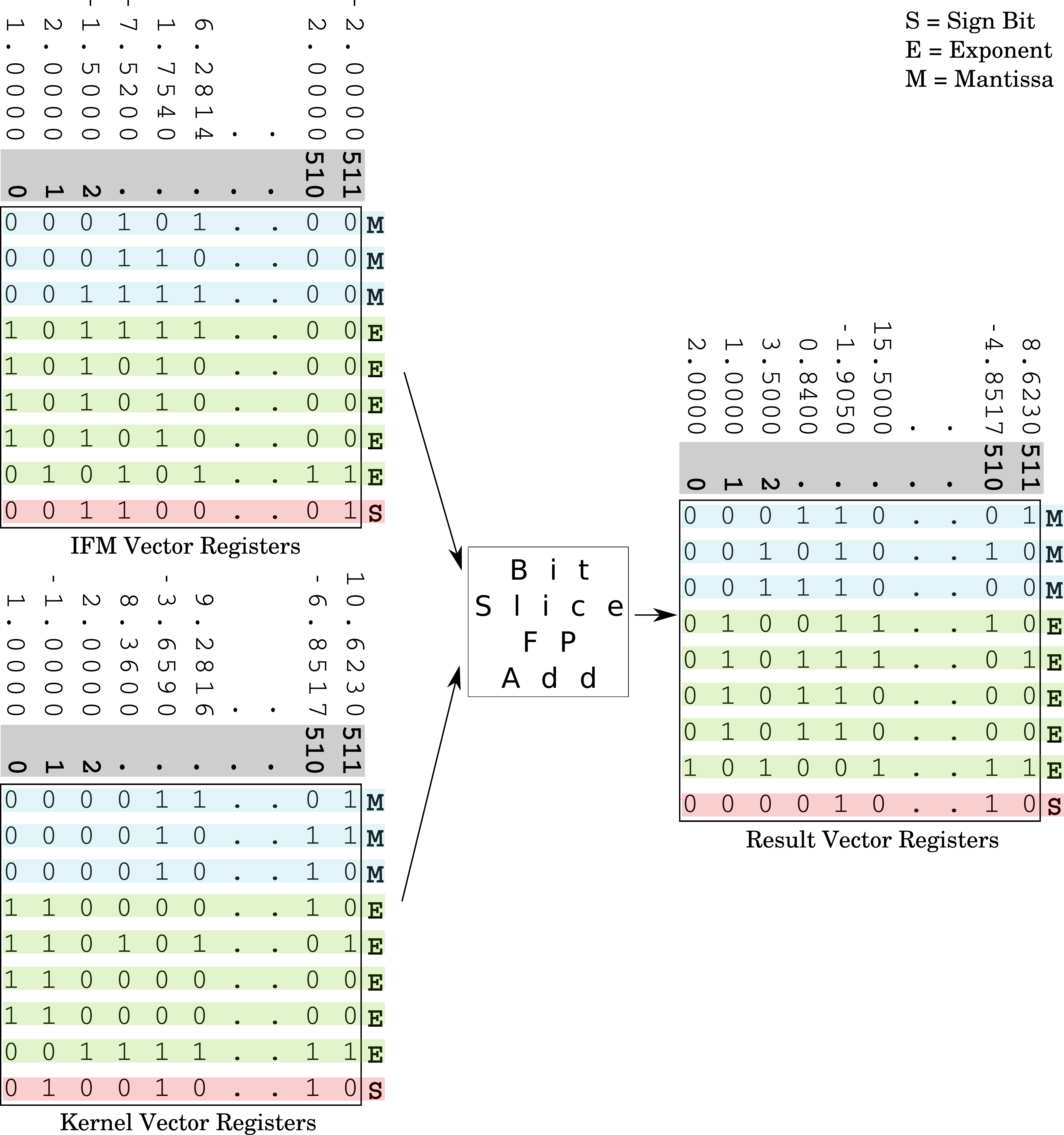}\label{fig:bitSliceOperation}}
  \caption{Bit-sliced Parallel FP Transformation and Operation.}
  \label{fig:bitSliceTransformAndOperation}
\end{figure}

\Gls{hobflops} exploits bitslice parallel operations to represent \gls{fp} numbers in a bitwise manner that are processed in parallel. For example, many 9-bit values are transformed to bitslice parallel representations, see Figure~\ref{fig:fp2bsfp}. A simple example of how nine registers of 512-bit bitslice parallel data are applied to a 512-bit wide bitslice parallel \gls{fp} adder is shown in Figure~\ref{fig:bitSliceOperation}. Each adder instruction has a throughput of around half a clock cycle (see Intel's Intrinsics Guide for details of the precise throughput of SSE, AVX2 and AVX512 logic Bitwise Operations and Arm's Intrinsics Reference for details of Arm Neon bitwise operational throughput). In this example, the adder's propagation delay is related to the number of instruction-level parallelism and associated load/store commands. The number of gates in the \gls{hobflops} adder or multiplier is dependent on the required \gls{hobflops} precision, see Table~\ref{tab:macTypes} for examples of \gls{hobflops} \gls{mac} precision, and Section~\ref{sec:evaluation} for associated \gls{hobflops} \gls{mac} gates counts and performance.

{\subsection{Design Flow}
  \label{subsec:designFlow}}
We investigate whether bitslice parallel logic can be optimized using hardware tools to reduce the hardware logic gate count or area, and subsequent lines of bitwise software operations. We use the industry-standard hardware \gls{asic} synthesizer, Cadence Genus, with our custom-designed logic cell libraries to synthesize and optimize the bitslice parallel \gls{fp} arithmetic. We use Yosys \gls{asic} synthesizer \cite{Yosys2013:Wolf} and ABC optimizer \cite{ABC2010:Brayton} to further optimize and topologically sort the netlist.

\begin{table}
\fontsize{8}{9}\selectfont
\centering
\caption{Comparison of Existing Custom FP.}
\label{tab:hobflopsRangePrecision}
\begin{tabular}{|l|l|l|l|}
\hline
\textbf{Sign} & \textbf{Exponent} & \textbf{Mantissa} & \textbf{Type \& Availability}   \\ \hline
1             & 4                 & 3                 & IEEE-FP8 in Software \cite{FpArithmeticStandard2019:IEEE} \\ \hline
1 & 5 & 2 & \begin{tabular}[c]{@{}l@{}}MS-FP8 in \gls{fpga} \cite{ProjectBrainwave2018:Chung,Brainwave2018:Fowers}\end{tabular}     \\ \hline
1 & 5 & 3 & \begin{tabular}[c]{@{}l@{}}MS-FP9 in \gls{fpga} \cite{ProjectBrainwave2018:Chung,Brainwave2018:Fowers}\end{tabular} \\ \hline
\end{tabular}%
\end{table}

\begin{table}
\fontsize{8}{9}\selectfont
\centering
\caption{\gls{hobflops} \gls{mac} Precision Types}
\label{tab:macTypes}
\begin{tabular}{|l|l|l|l|l|l|l|l|l|l|} 
\hline
\textbf{hobflops(IEEE)\textit{XX}} & \multicolumn{2}{l|}{\begin{tabular}[c]{@{}l@{}}\textbf{Inputs}\\\textbf{Bit Width}\end{tabular}} & \multicolumn{2}{l|}{\begin{tabular}[c]{@{}l@{}}\textbf{Outputs}\\\textbf{Bit Width}\end{tabular}} & \textbf{hobflops(IEEE)\textit{XXe}} & \multicolumn{2}{c|}{\begin{tabular}[c]{@{}c@{}}\textbf{Inputs}\\\textbf{ Bit Width} \end{tabular}} & \multicolumn{2}{c|}{\begin{tabular}[c]{@{}c@{}}\textbf{Outputs}\\\textbf{Bit Width} \end{tabular}}  \\ 
\hline
                                             & \textbf{Expo}         & \textbf{Mant}                                                            & \textbf{Expo}         & \textbf{Mant}                                                             & \textbf{ }                                    & \textbf{Expo}         & \textbf{Mant}                                                              & \textbf{Expo}         & \textbf{Mant}                                                               \\ 
\hline
Soft FP16 \cite{SoftFloat2002:Hauser}        & 5                     & 10                                                                       & 5                     & 10                                                                        &                                               &                       &                                                                            &                       &                                                                             \\ 
\hline
HOBFLOPSIEEE8                                & 4                     & 3                                                                        & 4                     & 4                                                                         & HOBFLOPSIEEE8e                                & 4                     & 3                                                                          & 4                     & 7                                                                           \\ 
\hline
HOBFLOPS8                                    & 5                     & 2                                                                        & 5                     & 3                                                                         & HOBFLOPS8e                                    & 5                     & 2                                                                          & 5                     & 5                                                                           \\ 
\hline
HOBFLOPS9                                    & 5                     & 3                                                                        & 5                     & 4                                                                         & HOBFLOPS9e                                    & 5                     & 3                                                                          & 5                     & 7                                                                           \\ 
\hline
\textbf{\textit{... (truncated)}}            & \textbf{\textit{...}} & \textbf{\textit{...}}                                                    & \textbf{\textit{...}} & \textbf{\textit{...}}                                                     & \textbf{\textit{... (truncated)}}             & \textbf{\textit{...}} & \textbf{\textit{...}}                                                      & \textbf{\textit{...}} & \textbf{\textit{...}}                                                       \\ 
\hline
HOBFLOPS16                                   & 5                     & 10                                                                       & 5                     & 11                                                                        & HOBFLOPS16e                                   & 5                     & 10                                                                         & 5                     & 21                                                                          \\
\hline
\end{tabular}
\end{table}

Taking inspiration from Microsoft's MS-\gls{fp}8 / MS-\gls{fp}9 \cite{ProjectBrainwave2018:Chung,Brainwave2018:Fowers} and Minifloat 8-bit that follows the IEEE-754 2019 specification, we create single-precision and extended-precision \gls{hobflops} adders and multipliers. For example, we create the single-precision \gls{hobflops}8 multiplier to take two 5-bit exponent, 2-bit mantissa, 1-bit sign inputs and produce a single 5-bit exponent and 3-bit mantissa and 1-bit sign output. We also create extended-precision \gls{hobflops}8e multiplier to take two 5-bit exponent, 2-bit mantissa, and 1-bit sign to produce a single 5-bit exponent, 5-bit extended mantissa and 1-bit sign output.

When using \gls{hobflops}, any quantization method may be employed, such as those investigated by Gong \etal \cite{CompressingDNNsUsingQuantization2014:Gong}. These quantized values need to be stored and computed during inference mode. Therefore we set \gls{flopoco} to the required range and precision. For comparison, Table~\ref{tab:hobflopsRangePrecision} shows the range and precision of IEEE-FP8 of MS-FP8 and MS-FP9, respectively, available in simulated software and \gls{fpga} hardware. These \gls{fp} solutions only support specific ranges and do not allow for arbitrary exponent and mantissa values.

Details of the evaluated \gls{hobflops} are in Table~\ref{tab:macTypes}, although, other arbitrary combinations of mantissa and exponent bit-widths are supported in the flow (see Figure~\ref{fig:workflow}).

\begin{figure*}
  \centering
  \includegraphics[width=0.4\linewidth]{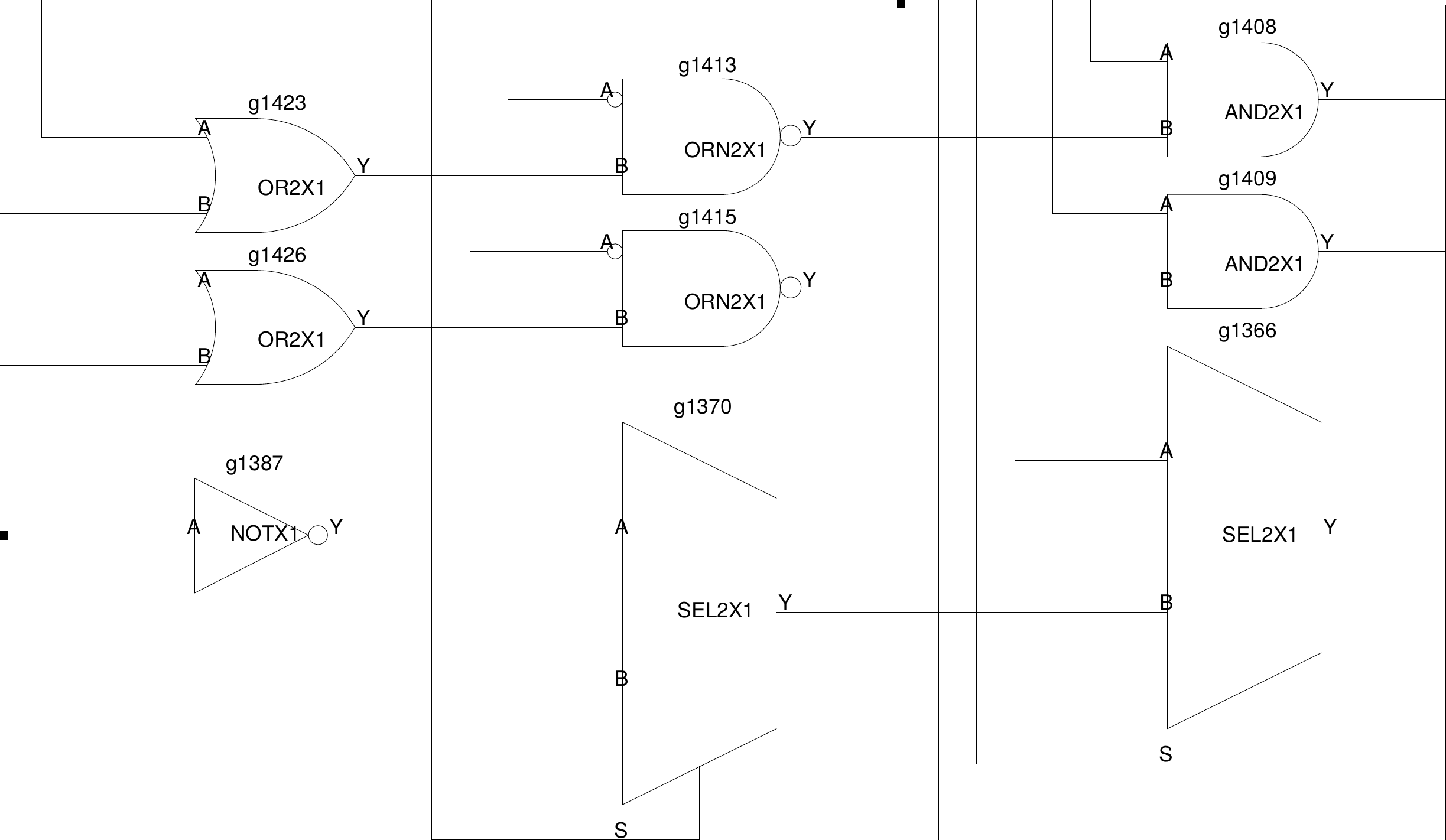}
  \caption{Zoomed Snippet of HOBFLOPS8 Multiplier Netlist with Arm Neon Cells.}
  \label{fig:hobflopsFp8MultNetlistSchematic}
\end{figure*}

\Gls{flopoco} \cite{GeneratingHighPerformanceCustomFpPipelines2009:Dinechin}, generates VHDL descriptions of \gls{fp} adders and multipliers of varying exponent and mantissa bit-widths for standard and extended precision \gls{hobflops} types, see Table~\ref{tab:macTypes}. As a performance baseline, we produce the 16- and 16e-bit IEEE-754 \gls{fp} versions of the multiplier and adder, for comparison against Berkeley's Soft\gls{fp}16. \Gls{flopoco} automatically produces the corresponding VHDL test benches and test vectors required to test the generated cores. \Gls{flopoco} has a slightly different way of encoding the \gls{fp} numbers when compared to the IEEE-754 2019 specification and does not support subnormal numbers. A \gls{flopoco} \gls{fp} number \cite{GeneratingHighPerformanceCustomFpPipelines2009:Dinechin} is a bit-vector consisting of the following 4 fields: 2-bit exception field (\textit{01} for normal numbers); a sign bit; an exponent field \textit{wE} bits wide; a mantissa (fractional) field \textit{wF} bits wide. The significand has an implicit leading \textit{1}, so the fraction field \textit{ff...ff} represents the significand \textit{1.ff...ff}.

We configure \gls{flopoco} to generate combinatorial plain \gls{rtl} VHDL cores with a 1MHz frequency, no pipelining, and no use of hard-macro \gls{fpga} multipliers or adders. These settings ensure that \gls{flopoco} generates reduced area rather than reduced latency multipliers and adders. We simulate the \gls{fp} multipliers and adders in a VHDL simulator with the corresponding \gls{flopoco} generated test bench to confirm that the quantized functionality is equivalent to IEEE-754 \gls{fp} multiplier and adder.

We create Synopsys Liberty standard cell libraries to support the target processor architecture. Cadence Genus (version 16.22-s033\textunderscore1) the industry-standard \gls{asic} synthesis tool, synthesizes the adder and multiplier VHDL cores with our standard cell libraries and configuration and small gate area optimisation script into a Verilog netlist of the logic gates. See Figure~\ref{fig:hobflopsFp8MultNetlistSchematic} for an example of the \gls{hobflops}8 multiplier logic produced by \gls{flopoco} when synthesized with our Arm Neon cell Library. Note how Genus has synthesized the design to include the 3-input SEL gate (multiplexer) supported by Arm Neon.

\Gls{hobflops} designs are combinatorial, so synthesis timing constraints are unnecessary. In the standard cell libraries Liberty file, we assign a value of 1.0 to \textit{cell area} and \textit{cell leakage power} of the cells. We configure the cell capacitance and timing values to zero. These values ensure the synthesizer assigns equal optimization priority to all gates and produces a netlist with the least number of logic gates rather than creating a netlist optimized for hardware timing propagation.

We further optimize the netlist using the open-source Yosys \gls{asic} synthesizer \cite{Yosys2013:Wolf} and ABC optimizer \cite{ABC2010:Brayton}. We use ABC's \textit{strash} command to transform the current network into an \gls{aig} by one-level structural hashing. We then use the \textit{refactor} function to iteratively collapse and refactor the levels of logic and area of the netlist. We configure Yosys to produce a topologically sorted Verilog netlist of gates. The topological sorting is required as Cadence Genus writes the netlist file in an output port to input port order, whereas the C/C++ compiler requires the converted netlist to have input to output ordering. We formally verify the topologically sorted netlist against the original netlist with Yosys \gls{sat}-solver. These netlists are re-simulated with the test bench used to simulate the \gls{flopoco} generated VHDL designs and compared for correlation.

\begin{figure*}
  \centering
  \includegraphics[width=0.9\linewidth]{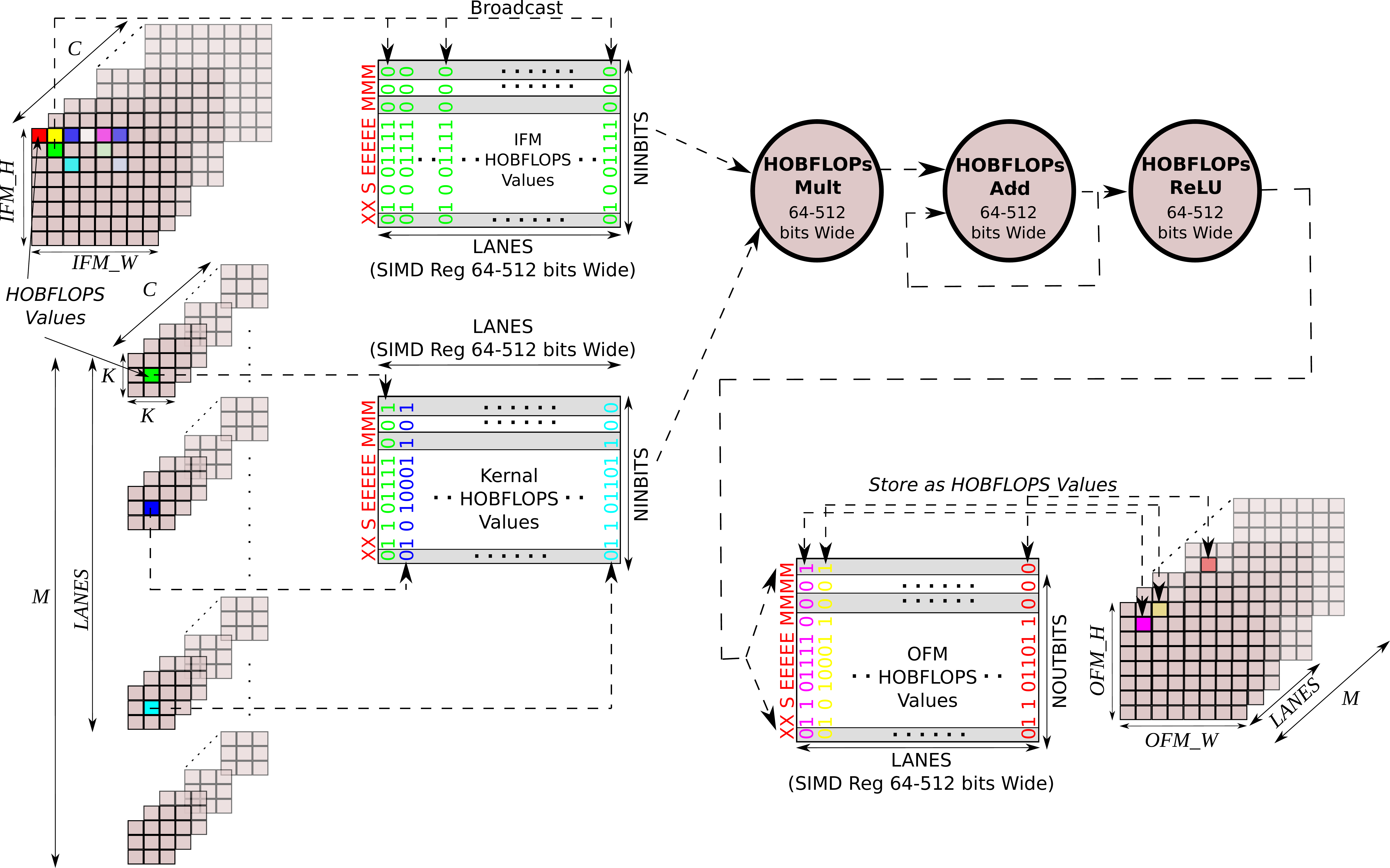}
  \caption{HOBFLOPS CNN Convolution (IFM and Kernel data pre-transformed to \gls{hobflops}, OFM remains in HOBFLOPs layout.)}
  \label{fig:hobflopsFlowDiagram}
\end{figure*}

\begin{lstlisting}[language={C++}, caption={Macros for AVX2 Cell Library Bitwise Operator Definitions}, label={lst:intelCellLibrary},float]
void AND2X1(u256 *A, u256 *B, u256 *Y) {
  *Y = _mm256_and_si256(*A, *B);}
void NOTX1(u256 *A, u256 *Y) {
  // Inverter could be implemented in many ways
  *Y = _mm256_xor_si256(*A, _mm256_set1_epi32(-1));}
void OR2X1(u256 *A, u256 *B, u256 *Y) {
  *Y = _mm256_or_si256(*A, *B);}
void XOR2X1(u256 *A, u256 *B, u256 *Y) {
  *Y = _mm256_xor_si256(*A, *B);}
void ANDNOT2X1(u256 *A, u256 *B, u256 *Y) {
  *Y = _mm256_andnot_si256(*A, *B);}
\end{lstlisting}

\begin{algorithm}
  \fontsize{8}{9}\selectfont
  \caption{\gls{hobflops} Code for a Simple AVX2 2-bit Binary Full Adder (unrolled by Synthesis) - 256 wide 2-bit adders in 12 Bitwise Operations.}
  \label{alg:hobFlops2BitAdderAvx2}
  \KwIn{ \textit{x[2]} of SIMD width}
  \KwIn{ \textit{y[2]} of SIMD width}
  \KwIn{ \textit{cin} of SIMD width}
  \KwOut{ \textit{HOBFLOPS} register \textit{sum[2]} of SIMD width}
  \KwOut{ \textit{HOBFLOPS} register \textit{cout} of SIMD width}
  XOR2X1(x[1], y[1], n\_5); \textit{// Wide XOR Operation} \\
  OR2X1(x[1], y[1], n\_2); \textit{// Wide OR Operation} \\
  OR2X1(cin, x[0], n\_1)\;
  AND2X1(x[1], y[1], n\_0); \textit{// Wide AND Operation} \\
  AND2X1(cin, x[0], n\_3)\;
  AND2X1(y[0], n\_1, n\_6)\;
  OR2X1(n\_3, n\_6, n\_8)\;
  AND2X1(n\_2, n\_8, n\_9)\;
  OR2X1(n\_0, n\_9, cout)\;
  XOR2X1(n\_5, n\_8, sum[1])\;
  XOR2X1(x[0], y[0], n\_4)\;
  XOR2X1(cin, n\_4, sum[0])\;
\end{algorithm}

\begin{algorithm}
  \fontsize{8}{9}\selectfont
  \caption{\gls{hobflops} Code for a Simple AVX512 2-bit Binary Full Adder - 512 wide 2-bit adders in 4 Bitwise Operations.}
  \label{alg:hobFlops2BitAdderAvx512}
  \KwIn{ \textit{x[2]} of SIMD width}
  \KwIn{ \textit{y[2]} of SIMD width}
  \KwIn{\textit{cin} of SIMD width}
  \KwOut{ \textit{HOBFLOPS} register \textit{sum[2]} of SIMD width}
  \KwOut{ \textit{HOBFLOPS} register \textit{cout} of SIMD width}
  LUT232X1(cin, y[0], x[0], n\_1); \textit{// $(B\&C)|A\&(B \wedge C)$} \\
  LUT232X1(x[1], n\_1, y[1], cout)\;
  LUT150X1(y[1], x[1], n\_1, sum[1]); \textit{// $A \wedge B \wedge C$} \\
  LUT150X1(y[0], x[0], cin, sum[0])\;
\end{algorithm}

Our domain-specific source to source generator translates the Verilog adder and multiplier netlists to Intel AVX2, AVX512, or Arm Neon bitwise operators with the correct types and pointers. Algorithm~\ref{alg:hobFlops2BitAdderAvx2} shows the \gls{hobflops} code for a simple 2-bit binary adder with carry, generated from 12 bitwise operations, referenced from the cell library of Listing~\ref{lst:intelCellLibrary}. A single input bit of the hardware multiplier becomes the corresponding architecture variable type \eg uint64 for a 64-bit processor, an \_\_mm256i type for an AVX2 processor, an \_\_mm512i type for an AVX512 processor, uint32x4\_t type for a Neon processor. Algorithm~\ref{alg:hobFlops2BitAdderAvx512} demonstrates the efficiency of the AVX512 implementation of the same simple 2-bit adder, generated from 4 bitwise operations.

A \gls{hobflops}8 multiplier targeted at the AVX2 processor, for example, is generated in 80 bitwise operations, and a \gls{hobflops}8 adder is generated in 319 bitwise operations. When targeted at the AVX512 processor, the \gls{hobflops}8 multiplier is generated in 53 bitwise operations, and the \gls{hobflops}8 adder is generated in 204 bitwise operations.

{\subsection{CNN Convolution with HOBFLOPS}
  \label{subsec:cnnConvolutionWithHobflops}}
We present a method for \gls{cnn} convolution with \gls{hobflops} arithmetic. We implement \gls{hobflops} \glspl{mac} in a \gls{cnn} convolution layer, where up to 90\% of the computation time is spent in a \gls{cnn} \cite{hwAccelCnn2010:Farabet}. We compare \gls{hobflops} \gls{mac} performance to IEEE \gls{fp} \gls{mac} and to Berkeley's Soft\gls{fp}16 \textit{MulAdd} function \cite{SoftFloat2002:Hauser}.

Figure~\ref{fig:hobflopsFlowDiagram} shows \gls{hobflops} \gls{ifm} and kernel values convolved and stored in the \gls{ofm}. To reduce cache misses, we tile the \gls{ifm} $H \times W \times C$ dimensions, which for \textit{Conv dw / s2} of MobileNets \gls{cnn} is $14 \times 14 \times 512 = 100,352$ elements of \gls{hobflops} \gls{ifm} values. We tile the $M$ kernel values by $LANES \times NINBITS$, where $LANES$ corresponds to the target architecture registers bit-width, \eg 512-lanes corresponds to AVX512 512-bit wide register. The $LANES \times NINBITS$ tiles of binary values are transformed to $NINBITS$ of \textit{SIMD width} values, where \textit{SIMD width} correspond to the target architecture register width, \eg uint64 type for a 64-bit processor architecture, \_\_mm512i type for AVX512 processor.

We broadcast the \gls{ifm} channel tile of $NINBITS$ across the corresponding channel of all the kernels tiles of $NINBITS$ to convolve image and kernel values using \gls{hobflops} multipliers, adders and \gls{relu} activation function. The resultant convolution \textit{SIMD width} values of $NOUTBITS$ wide are stored in corresponding location tiles in the \gls{ofm}. The \gls{hobflops} \gls{ifm} and kernel layout for single-precision, as defined by \gls{flopoco} is:\\
\indent NINBITS=EXC+SIGN+EXPO\_IN+MANT\_IN\\
The \gls{hobflops} \gls{ofm} layout for single-precision is:\\
\indent NOUTBITS=EXC+SIGN+EXPO\_IN+MANT\_IN+1\\
and for extended-precision:\\
\indent NOUTBITS=EXC+SIGN+EXPO\_IN+(2$\times$MANT\_IN)+1\\
For example, \gls{hobflops}9, as can be seen in Table~\ref{tab:macTypes}, the input layout $NINBITS$ has 2-bit exception $EXC$, 1-bit sign $SIGN$, 5-bit exponent $EXPO\_IN$, 3-bit mantissa $MANT\_IN$, which added comes to 11-bits. The single-precision $NOUTBITS$ would be 12-bits (essentially $NINBITS+1$). The extended-precision $NOUTBITS$ would be 15-bits.

If \gls{hobflops} is implemented in a multi-layer \gls{cnn}, the data between each layer could remain in \gls{hobflops} format until the last convolution layer. The \gls{ofm} at the last convolutional layer could be transformed from \gls{hobflops} values to floats resulting in the transformation overhead only occurring at the first and last convolutional layers of the \gls{cnn} model. An additional pooling layer could be developed in the \gls{hobflops} format, for the interface between the last convolutional layer and the fully connected layer of MobileNets, not done in this work. 

We repeat the above for \glspl{mac} up to \gls{hobflops}16e on each processor architecture up to the 512-bit wide AVX512 registers.

\section{Evaluation}\label{sec:evaluation}
We implement each of the 8- to 16e-bit \gls{hobflops} multipliers and adders in a convolution layer of the MobileNets \gls{cnn} \cite{Mobilenets2017:Howard}. We use layer \textit{Conv dw / s2} of MobileNets as it has a high number of channels $C$ and kernels $M$, perfect for demonstrations of high-dimensional parallelized \gls{mac} operations. We compare the \gls{hobflops}16 multipliers and adders round-to-nearest-ties-to-even and round-towards-zero modes performance to Berkeley's Soft\gls{fp}16 \textit{MulAdd} rounding near\_even and round min modes \cite{SoftFloat2002:Hauser}. This 16-bit \gls{fp} comparison acts as a baseline as Soft \gls{fp}8 is not supported by Berkeley's emulation tool.

We target 32-bit to 512-bit registers for AVX2 and AVX512 processors and target 32- and 128-bit registers for the Cortex-A15 processor. We implement 32- to 512-lanes of \gls{hobflops} multipliers and adders and capture each of the AVX2 32-, 64-, 128- and 256-bit, AVX512 32-, 64-, 128-, 256 and 512-bit, and Cortex-A16 32-, 64- and 128-bit results. 

\noindent Three machine types are used to test the \gls{hobflops} \gls{mac}:
\begin{itemize}
  \item Arm Cortex-A15 Neon embedded development kit, containing an ARMv7 rev 3 (v7l) CPU at 2GHz and 2GB RAM;
  \item Intel Core-i7 PC, containing Intel Core-i7 8700K CPU at 3.7GHz and 32GB RAM;
  \item Intel Xeon Gold server PC, containing Intel Xeon Gold 5120 at 2.2GHz and 256GB RAM. 
\end{itemize}

\noindent Our cell libraries model-specific cells, see Table~\ref{tab:gatesSupported}. We omit the \acrfull{bic} of the Arm Neon in our cell library. Inclusion of \gls{bic} prevents the synthesis tool, Cadence Genus, from optimizing the netlist with SEL (multiplexer) units, leading to a less area efficient netlist.

To further decrease area and increase performance, we produce round-towards-zero versions of the \gls{hobflops}8--\gls{hobflops}16e adders as the rounding can be dealt with at the end of the layer in the activation function, assuming the non-rounded part of the \gls{fp} value is retained through to the end of the layer.

The \glspl{mac} per second of an average of 1000 iterations of a \gls{hobflops} adders and multipliers are captured and compared. We use the GNU G++ compiler to optimize the code for the underlying target microprocessor architecture and numbers of registers. We compile the \gls{hobflops} \gls{cnn} code (see Figure~\ref{fig:hobflopsFlowDiagram}) to include our \gls{hobflops} adders and multipliers, and our cell library (\eg Listing~\ref{lst:intelCellLibrary} for AVX2) with G++ (version 8.2.1 20181127 on Arm, version 9.2.0 on Intel AVX2 and version 6.3.0 20170516 on Intel AVX512 machines). We target C++ version 17 and using \textit{-march=native,~-mtune=native,~-fPIC,~-O3} compiler switches with $-msse$ for SSE devices and \textit{-mavx2} for AVX2 devices. When targeting an Intel AVX512 architecture we use the the \textit{-march=skylake-avx512,~-mtune=skylake-avx512,~-mavx512f,~-fPIC,~-O3} switches. When targeting an Arm Neon device we use \textit{-march=native,~-mtune=native,~-fPIC,~-O3,~-mfpu=neon} to exploit the use of Neon registers.

After the G++ compilation, we inspect the assembler object dump. Within the multiplier and adder units, we find an almost one-to-one correlation of logic bitwise operations in the assembler related to the gates modeled in the cell libraries, with additional loads/stores where the compiler has seen fit to implement.

{\subsection{Arm Cortex-A15}
  \label{subsec:armCortexA15}} 
We configure an Arm Cortex-A15 Development kit with 2GB RAM, ARCH Linux version 4.14.107-1-ARCH installed, and fix the processor frequency at 2GHz. We run tests for 32-, 64- and 128-lanes and capture performance. We use \textit{taskset} to lock the process to a core of the machine for measurement consistency.

\begin{figure}
  \centering
  \subfloat[Arm Neon HOBFLOPS8-16e \glspl{mac} Round-to-Nearest-Ties-To-Even Throughput - \textbf{higher is better}.]{\includegraphics[width=7.8cm]{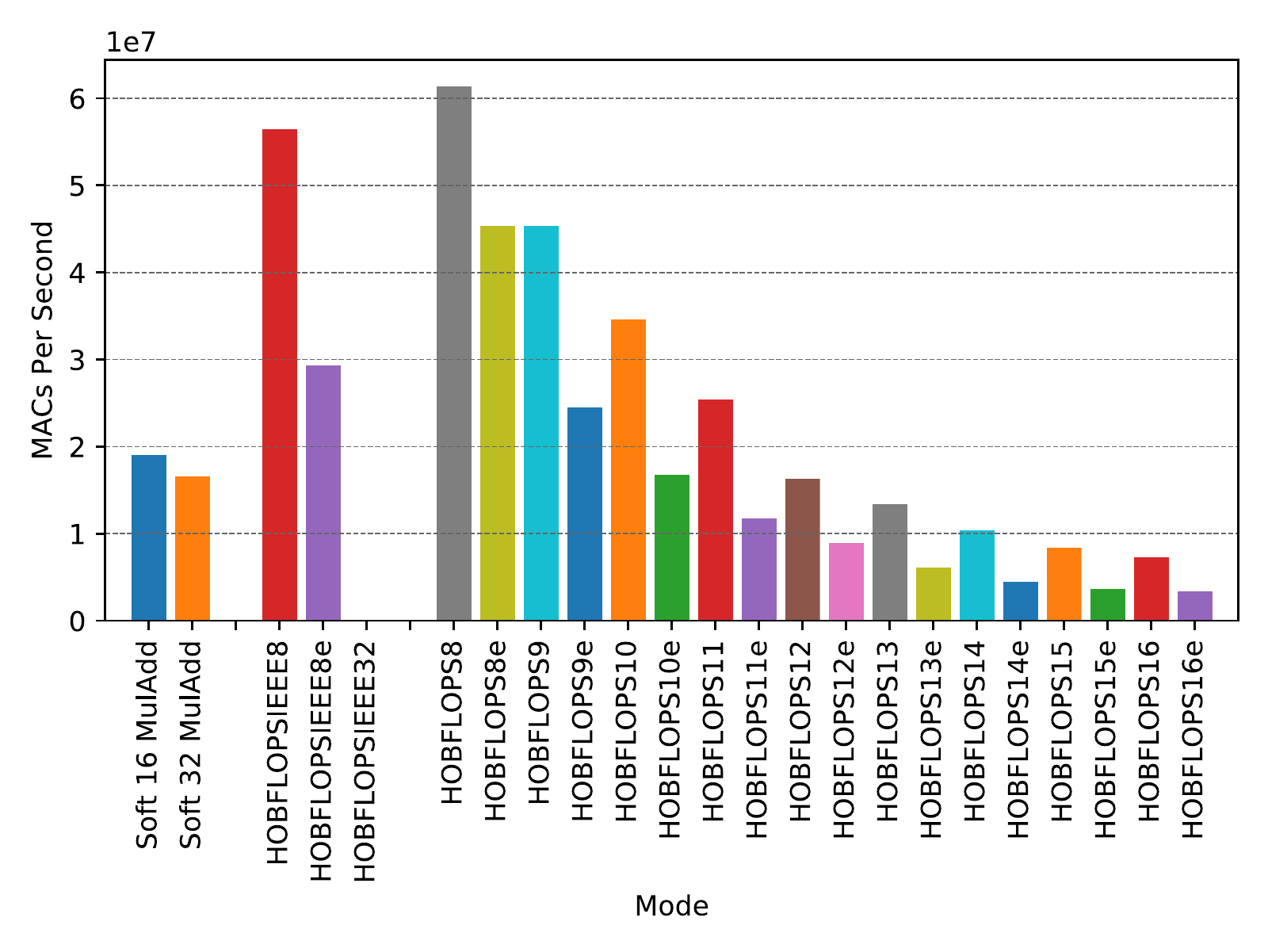}\label{fig:armNeonPerfR2Nearest}}\quad
  \subfloat[Arm Neon HOBFLOPS8-16e \glspl{mac} Round-Towards-Zero Throughput - \textbf{higher is better}.]{\includegraphics[width=7.8cm]{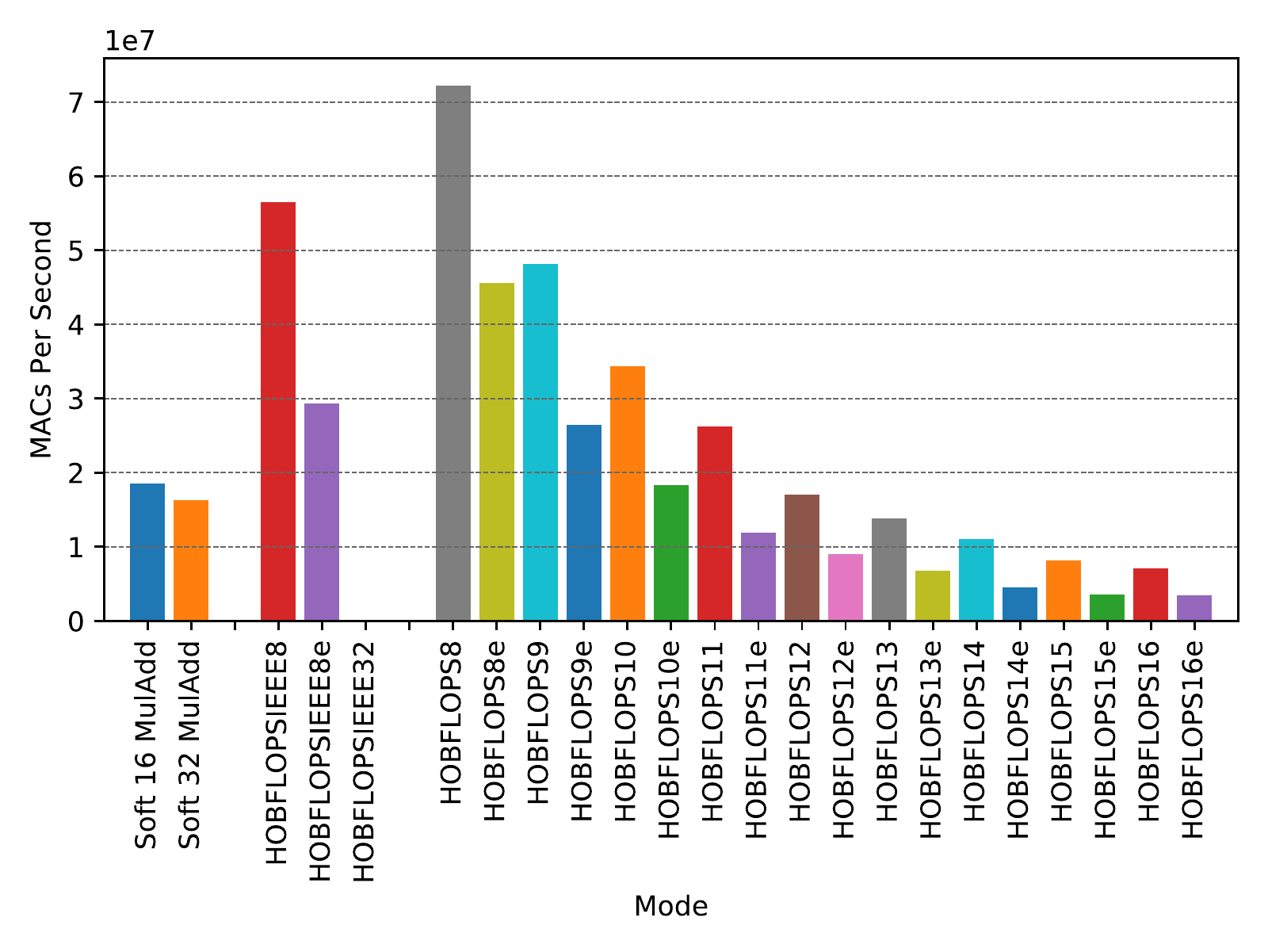}\label{fig:armNeonPerfR2Zero}}
  \caption{Arm Neon HOBFLOPS8-16e Performance}
  \label{fig:armNeonPerformance}
\end{figure}

\begin{figure}
  \centering
  \subfloat[Arm Neon HOBFLOPS8-16e \gls{mac} Gate Count: Round To Nearest, Ties To Even - \textbf{lower is better}.]{\includegraphics[width=7.8cm]{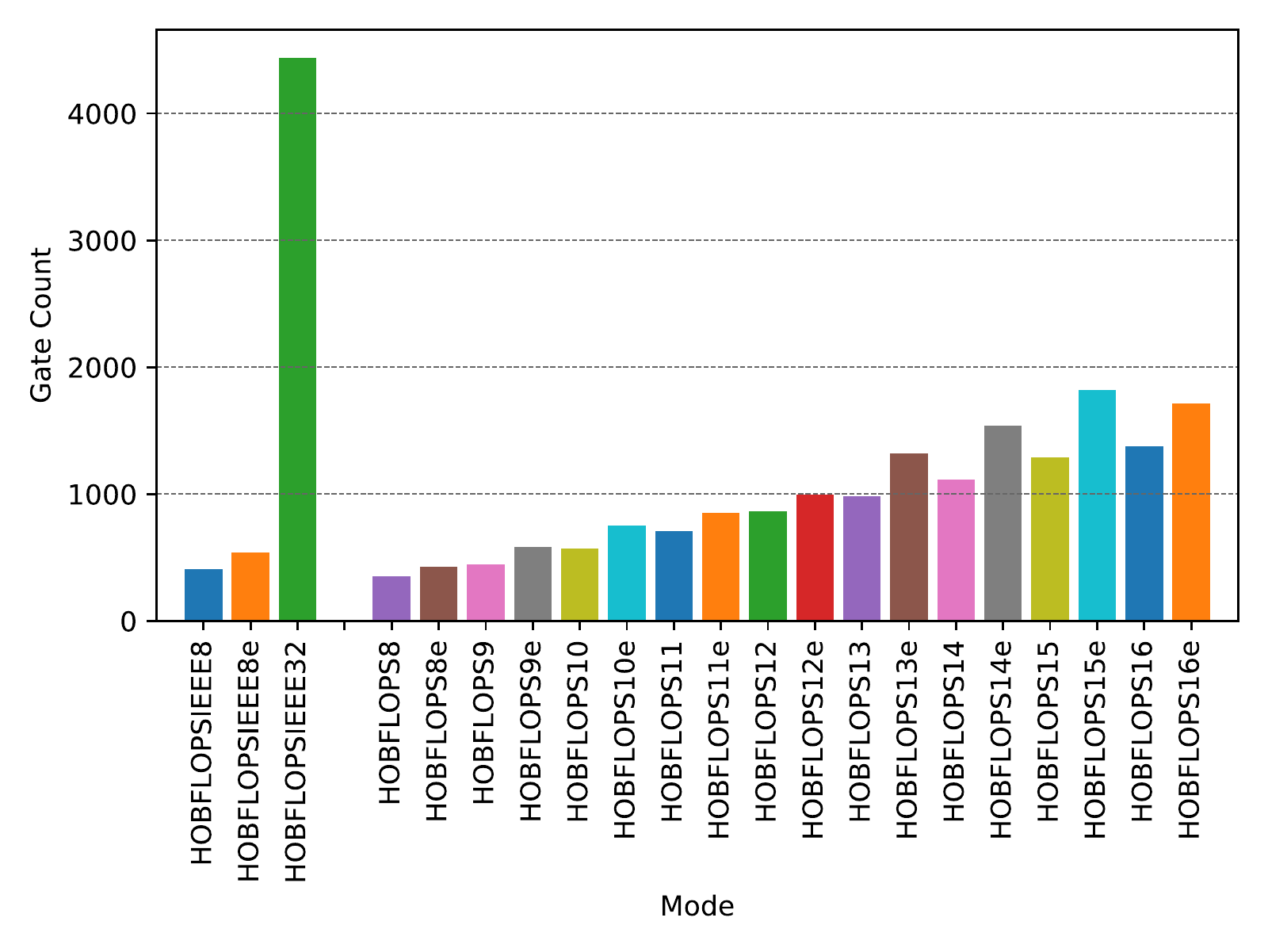}\label{fig:armNeonGateCountR2Nearest}}\quad
  \subfloat[Arm Neon HOBFLOPS8-16e \gls{mac} Gate Count: Round Towards Zero - \textbf{lower is better}.]{\includegraphics[width=7.8cm]{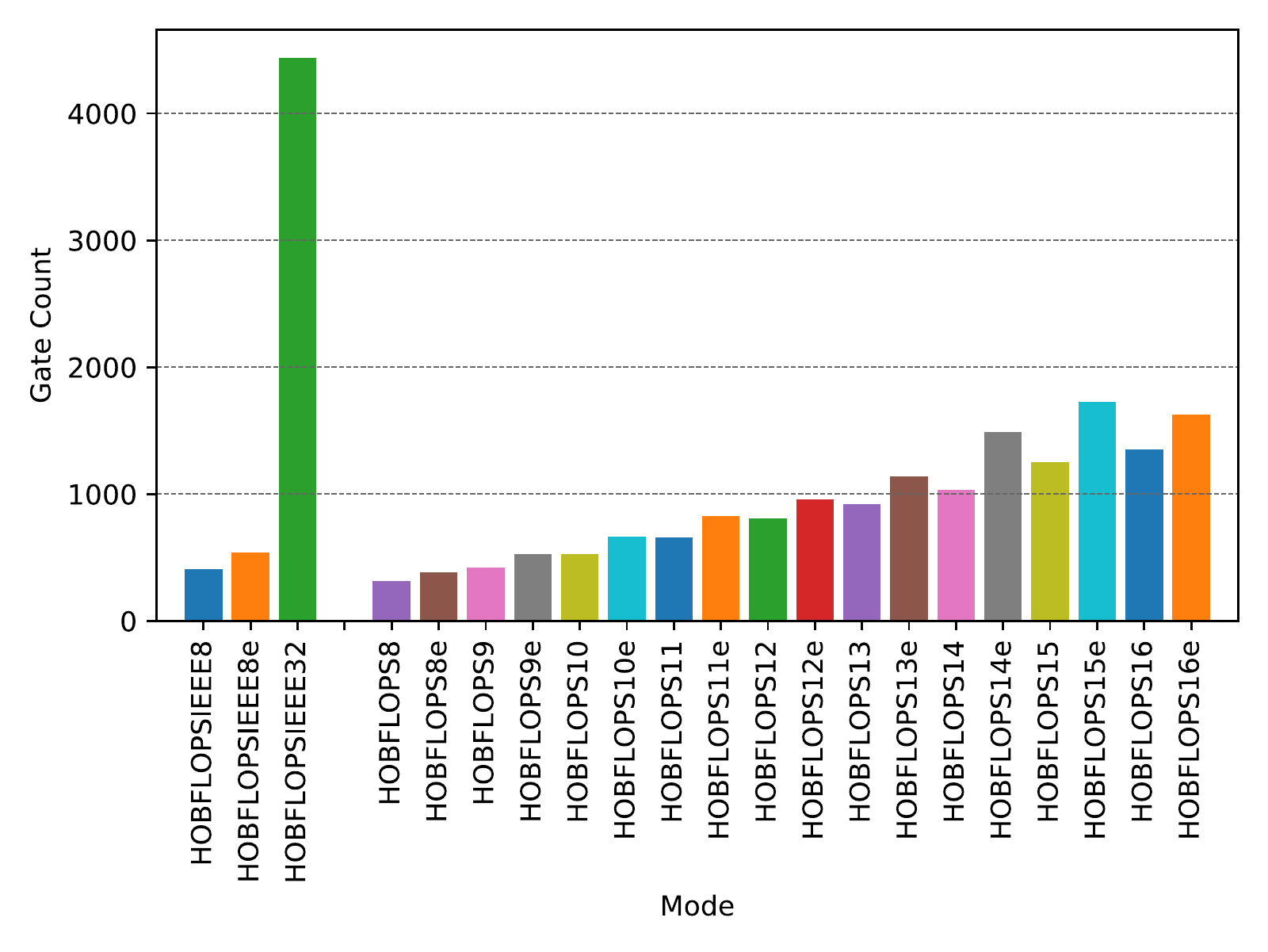}\label{fig:armNeonGateCountR2Zero}}
  \caption{Arm Neon HOBFLOPS8-16e Gate Count }
  \label{fig:armNeonGateCount}
\end{figure}

Figure~\ref{fig:armNeonPerfR2Nearest} shows 128-lane round-to-nearest-ties-to-even performance for all arbitrary-precision \gls{hobflops} \gls{fp} between 8- and 16e-bits, IEEE 8- and 32-bit equivalents and Berkeley's Soft\gls{fp} versions. \Gls{hobflops}16 round-to-nearest-ties-to-even achieves approximately half the performance of Soft\gls{fp}16 \textit{MulAdd} rounding near\_even mode on Arm Neon. However, \gls{hobflops} offers arbitrary precision mantissa and exponent \gls{fp} between 8- and 16-bits, outperforming Soft\gls{fp}16 between \gls{hobflops}8 and \gls{hobflops}11 bits.

Similarly, \gls{hobflops}16 round-towards-zero version shown in Figure~\ref{fig:armNeonPerfR2Zero} demonstrates a slight improvement in performance compared to Berkeley's Soft\gls{fp}16 \textit{MulAdd} rounding min mode. Figure~\ref{fig:armNeonPerfR2Zero} also shows \gls{hobflops} round-towards-zero versions have an increased performance when compared to \gls{hobflops}16 round-to-nearest-ties-to-even. 

\Gls{hobflops} appears to exhibit a fluctuation around \gls{hobflops}8e and \gls{hobflops}9 between Figure~\ref{fig:armNeonPerfR2Nearest} and Figure~\ref{fig:armNeonPerfR2Zero}. While there is 1-bit more in the input mantissa of \gls{hobflops}9 compared to \gls{hobflops}8e, which leads to \gls{hobflops}9 containing larger adder/accumulators, Figure~\ref{fig:armNeonPerfR2Zero} shows the round-towards-zero \gls{hobflops}9 functionality almost counter-intuitively exhibiting slightly greater throughput than \gls{hobflops}8e. The greater throughput of the round-towards-zero \gls{hobflops}9 is due to the lack of rounding adder, reduced gate area and latency.

The low bit-width and thus reduced hardware synthesis gate count or area as seen in Figure~\ref{fig:armNeonGateCountR2Nearest} and Figure~\ref{fig:armNeonGateCountR2Zero} would benefit memory storage and bandwidth within the embedded system allowing for reduced energy consumption, however, energy consumption is not measured here.

{\subsection{Intel AVX2}
  \label{subsec:intelAvx2}} 
We configure an Intel Core i7-8700K desktop machine with 32GB RAM, and ARCH Linux 5.3.4-arch1-1 installed. For consistency of performance measurements of various \gls{hobflops} configurations, within the BIOS we disable:
\begin{itemize}
  \item Intel's SpeedStep (\ie prevent the CPU performance from ramping up and down);
  \item Multi-threading (\ie do not split the program into separate threads);
  \item TurboBoost (\ie keep all processor cores running at the same frequency);
  \item Hyperthreading Control (\ie keep one program on one processor core);
  \item C-States control (\ie prevent power saving from ramping down the core clock frequency).
\end{itemize}
We alter GRUB's configuration so \textit{intel\_pstate} (\ie lock the processor core clock frequency) and \textit{intel\_cstate} are disabled on both \textit{GRUB\_CMDLINE\_LINUX} and \textit{GRUB\_CMDLINE\_LINUX\_DEFAULT}. This BIOS and Linux Kernel configuration ensures the processor frequency is fixed at 4.6GHz, no power-saving, and each \gls{hobflops} instance running at full performance on a single thread and single CPU core. When executing the compiled code, \textit{taskset} is used to lock the process to a single core of the CPU. These configurations allow a reproducible comparison of timing performance of each \gls{hobflops} configuration against Berkeley's Soft\gls{fp}16.

\begin{figure}
  \centering
  \subfloat[Intel AVX2 HOBFLOPS8-16e \glspl{mac} Round-to-Nearest-Ties-To-Even Throughput - \textbf{higher is better}.]{\includegraphics[width=7.8cm]{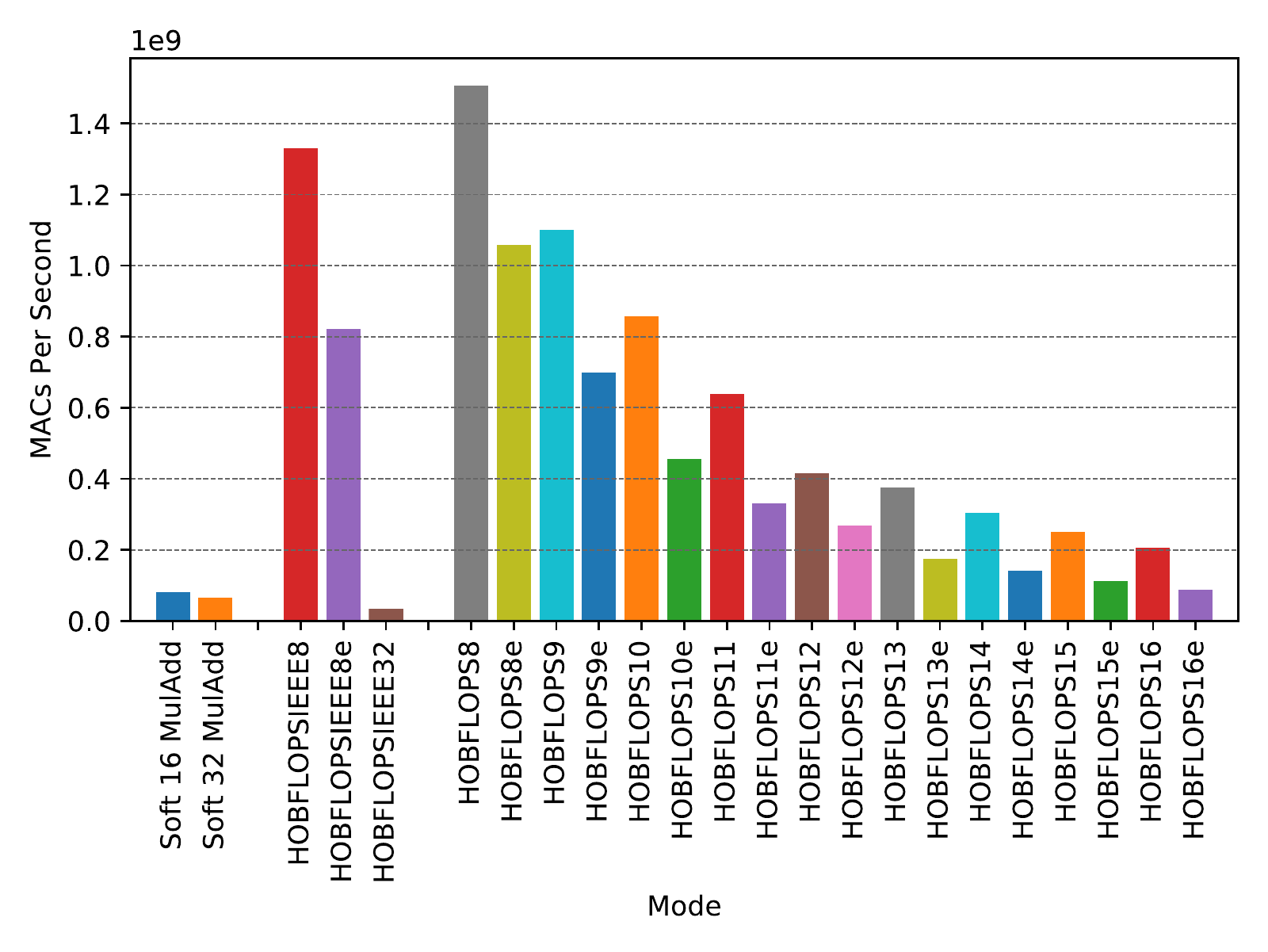}\label{fig:intelAvx2PerfR2Nearest}}\quad
  \subfloat[Intel AVX2 HOBFLOPS8-16e \gls{mac} Gate Count: Round To Nearest, Ties To Even - \textbf{lower is better}.]{\includegraphics[width=7.8cm]{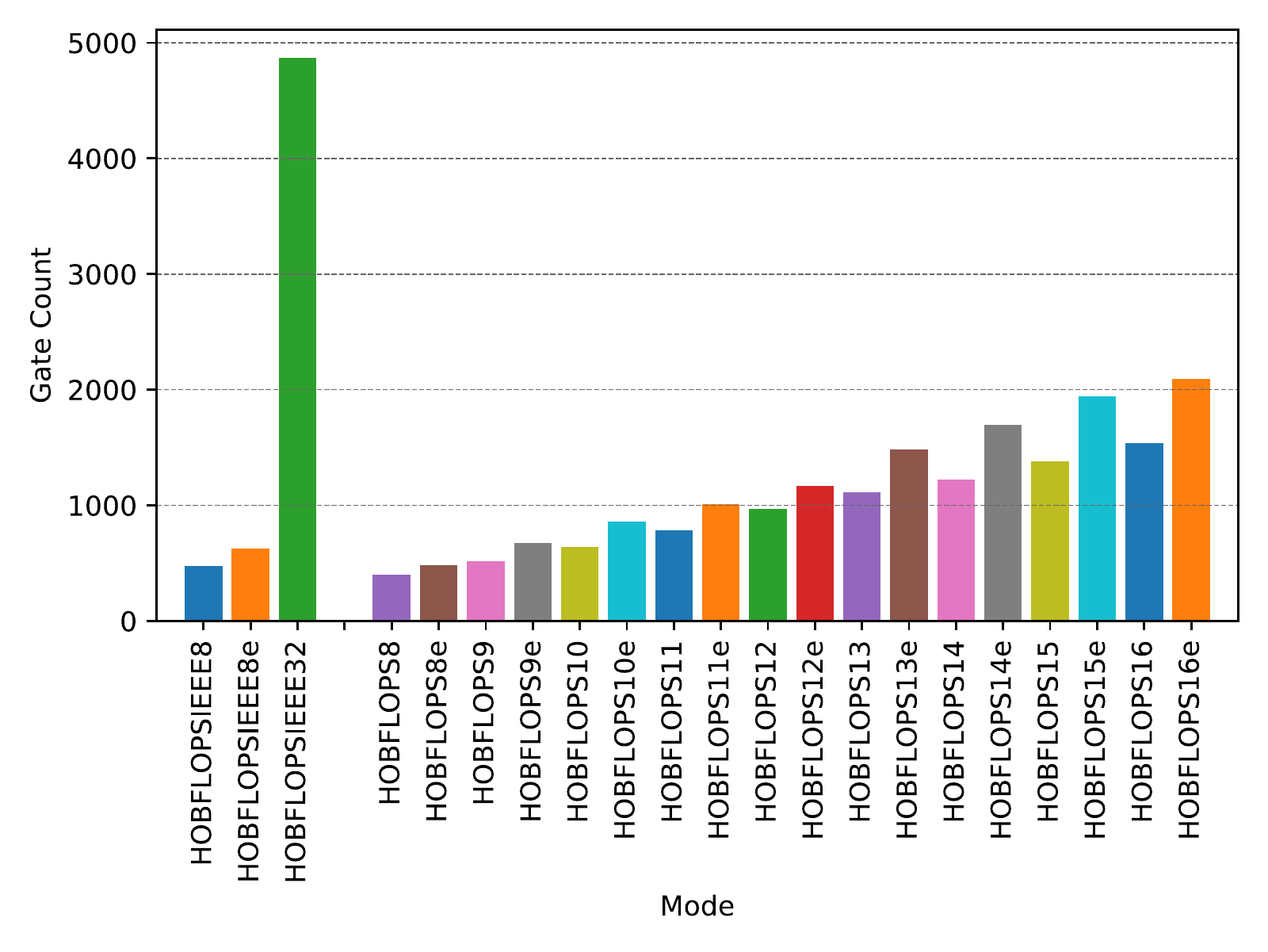}\label{fig:intelAvx2GateCountR2Nearest}}
  \caption{Intel AVX2 HOBFLOPS8-16e Performance and Gate Count}
  \label{fig:intelAvx2PerformanceAndGateCount}
\end{figure}

We run tests for 32-, 64-, 128- and 256-lanes and capture performance. Figure~\ref{fig:intelAvx2PerfR2Nearest} shows 256-lane round-to-nearest-ties-to-even results for all arbitrary-precision \gls{hobflops} \gls{fp} between 8- and 16e-bits, IEEE 8- and 32-bit equivalents and Berkeley's Soft\gls{fp} versions. \Gls{hobflops}16 performs over $2.5\times$ higher \glspl{mac}/second when compared to Berkeley's Soft\gls{fp}16 \textit{MulAdd} rounding near\_even mode. The round-towards-zero version of \gls{hobflops}16 performs at around $2.7\times$ higher \glspl{mac}/second when compared to Berkeley's Soft\gls{fp}16 \textit{MulAdd} rounding min mode. In fact, \gls{hobflops} outperforms Soft\gls{fp}16 for all versions of between \gls{hobflops}8 and \gls{hobflops}16e for both round-to-nearest-ties-to-even and round-towards-zero rounding modes.

\Gls{hobflops}8 performance gain is due to reduced synthesis area of the \gls{hobflops} units as seen in Figure~\ref{fig:intelAvx2GateCountR2Nearest}. Again, this reduction in area, also seen for round-towards-zero, is key to reduced \gls{simd} bitwise operations being created for the \gls{hobflops} \glspl{mac} and therefore reduced latency through the software \gls{hobflops} \glspl{mac}.

{\subsection{Intel AVX512}
  \label{subsec:intelAvx512}} 
We configure an Intel Xeon Gold 5120 server with 256GB RAM, and Debian Linux 4.9.189-3+deb9u2. This shared server-grade machine BIOS or clock could not be changed as done for the AVX2-based machine. However, \textit{taskset} is used to lock the process to a single CPU core.

\begin{figure}
  \centering
  \subfloat[Intel AVX512 HOBFLOPS8-16e \glspl{mac} Round-to-Nearest-Ties-To-Even Throughput - \textbf{higher is better}.]{\includegraphics[width=7.8cm]{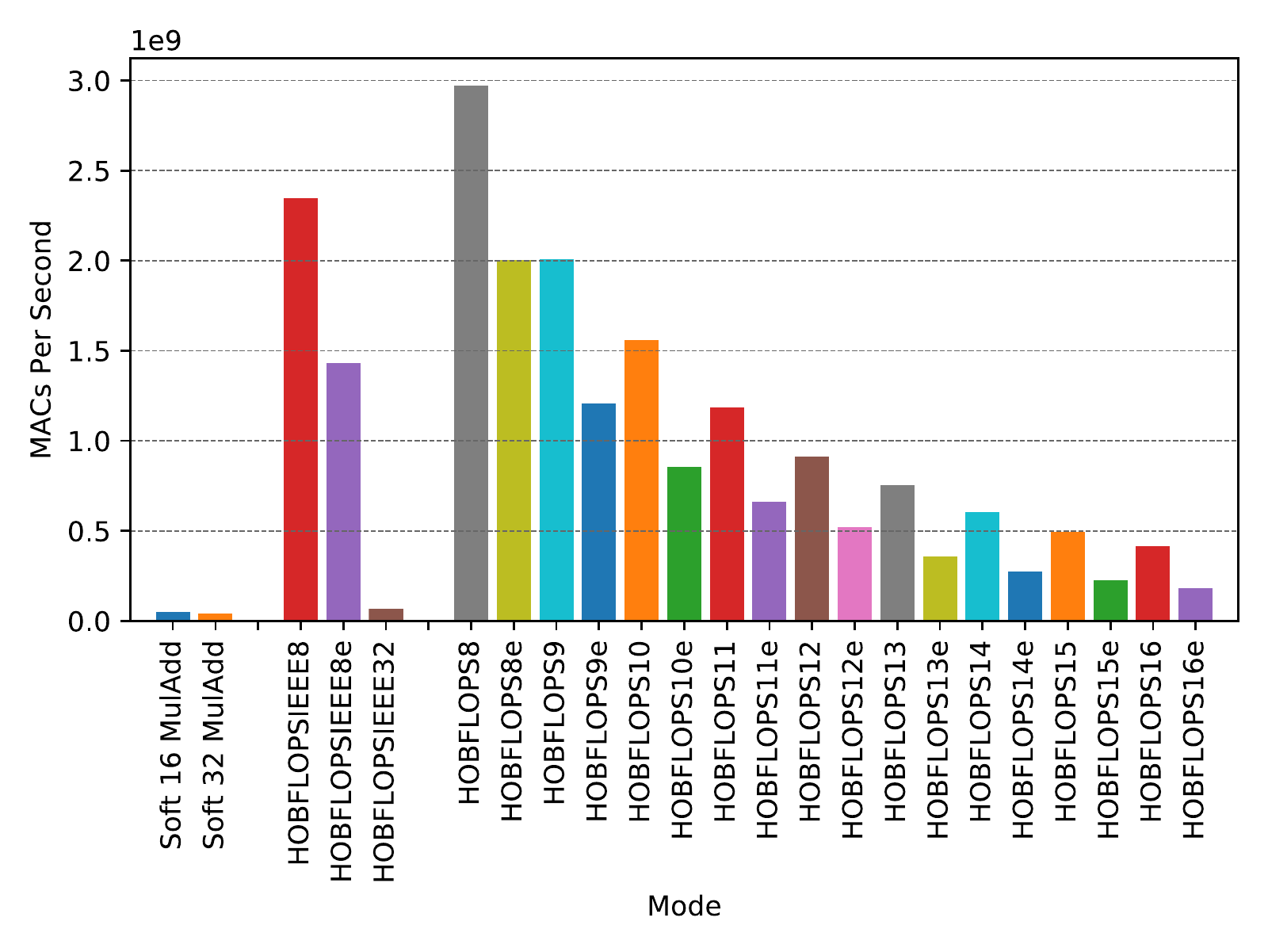}\label{fig:intelAvx512PerfR2Nearest}}\quad
  \subfloat[Intel AVX512 HOBFLOPS8-16e \gls{mac} Gate Count: Round To Nearest, Ties To Even - \textbf{lower is better}.]{\includegraphics[width=7.8cm]{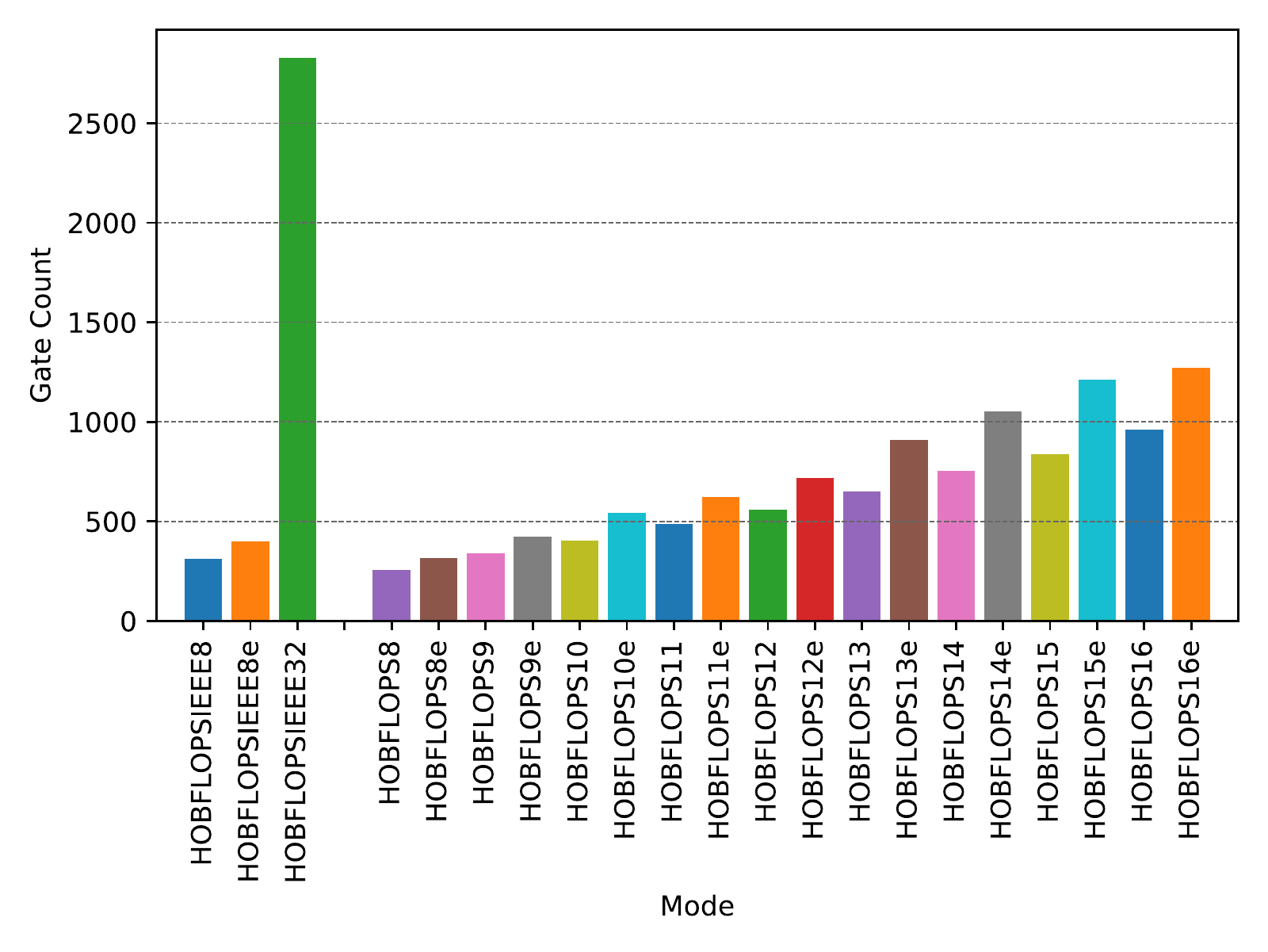}\label{fig:intelAvx512GateCountR2Nearest}}
  \caption{Intel AVX2 HOBFLOPS8-16e Performance and Gate Count}
  \label{fig:intelAvx512PerformanceAndGateCount}
\end{figure}

We run tests for 32-, 64-, 128-, 256- and 512-lanes. Figure~\ref{fig:intelAvx512PerfR2Nearest} captures 512-lane round-to-nearest-ties-to-even results for \gls{hobflops} \gls{fp} between 8- and 16e-bits, IEEE 8- and 32-bit equivalents and Berkeley's Soft\gls{fp} versions. \Gls{hobflops}16 performs with $8.2\times$ greater \glspl{mac} throughput than Soft\gls{fp}16 \textit{MulAdd} rounding near\_even mode. For the 512-lane round-towards-zero results, \gls{hobflops}16 performs at $8.4\times$ the \glspl{mac} throughput of Soft\gls{fp}16 \textit{MulAdd} rounding min mode. \Gls{hobflops} outperforms Soft\gls{fp}16 for \gls{hobflops}8 and \gls{hobflops}16e for both round-to-nearest-ties-to-even and round-towards-zero. \Gls{hobflops}9 performs at approximately 2 billion MACs/second, around $5\times$ the performance of \gls{hobflops}16.

\Gls{hobflops} performance is due to \gls{hobflops} lower hardware synthesis area, which when the netlists are converted to software bitwise operations, translates to fewer \gls{simd} 3-input ternary logic \gls{lut} instructions in the \glspl{mac}. As seen in Figure~\ref{fig:intelAvx512GateCountR2Nearest}, \gls{hobflops}16 area on the AVX512 platform is 38\% smaller than the \gls{hobflops}16 area on AVX2. A further slight performance boost is seen for round-towards-zero.

\section{Conclusion}\label{sec:conclusion}
We propose \gls{hobflops}, a method of generating fast custom-precision emulated bitslice parallel software \gls{fp} arithmetic using a hardware design flow, our cell libraries and custom code-generator. We generate efficient software-emulated \gls{fp} operators with an arbitrary precision mantissa and exponent. \gls{hobflops} offers \gls{fp} with custom range and precision, useful for \gls{fp} \gls{cnn} acceleration where memory storage and bandwidth are limited.

We experiment with large numbers of channels and kernels in \gls{cnn} convolution. When \gls{hobflops}16 \gls{mac} is implemented in the convolution layer on Arm Neon and Intel AVX2 and AVX512 processors and compared to Berkeley's Soft\gls{fp}16 \textit{MulAdd} \gls{fp} emulation, \gls{hobflops} achieves approximately $0.5\times$, $2.5\times$ and $8\times$ the performance of Soft\gls{fp}16 respectively. We show \eg \gls{hobflops}9 performs at approximately 2 billion \glspl{mac}/second on an AVX512, around $5\times$ the performance of \gls{hobflops}16 and approximately 45 million \glspl{mac}/second on Arm Neon processor around $6\times$ that of \gls{hobflops}16.

The performance gains are due to the optimized hardware synthesis area of the \glspl{mac}, which translates to fewer bitwise operations. Additionally, the bitslice parallelism of the very wide vectorization of the \glspl{mac} of \gls{cnn} convolution contributes to the performance boost. While we show results for 8- and 16-bit with a fixed exponent, \gls{hobflops} supports the emulation of any required precision of \gls{fp} arithmetic at any bit-width of mantissa or exponent, \eg \gls{fp}9 containing a 1-bit sign, 5-bit exponent and 3-bit mantissa, or \gls{fp}11 containing a 1-bit sign, 5-bit exponent and 5-bit mantissa, not supported with other software \gls{fp} emulation methods.

\bibliographystyle{unsrt}  
\bibliography{bibliography}  

\end{document}